\documentclass[12pt]{article}
\pdfoutput=1
\usepackage[sumlimits,intlimits,namelimits]{amsmath}
\usepackage{amssymb}
\usepackage[english]{babel}
\usepackage{bbm}
\usepackage{parskip}
\usepackage[para]{footmisc}
\usepackage[margin=1cm]{caption}

\usepackage{graphicx}          

\usepackage{cite}

\usepackage[makeroom]{cancel}
\usepackage[usenames,dvipsnames]{color}
\usepackage[dvipsnames]{xcolor}

\usepackage[bordercolor=black,backgroundcolor=yellow,linecolor=black,
textsize=scriptsize,shadow]{todonotes}

\newcommand{\arctanh}{\text{arctanh}}

\newcommand{\mC}{\mathcal{C}}
\newcommand{\mD}{\mathcal{D}}

\newcommand{\mS}{\mathcal{S}}
\newcommand{\mL}{\mathcal{L}}

\newcommand{\mK}{\mathcal{K}}
\newcommand{\mN}{\mathcal{N}}
\newcommand{\mO}{\mathcal{O}}

\newcommand{\totd}{\text{d}}

\newcommand{\mP}{\mathcal{P}}
\newcommand{\mQ}{\mathcal{Q}}
\newcommand{\mV}{\mathcal{V}}

\newcommand{\Poincare}{Poincar\'e }

\newcommand{\tcw}{\textcolor{white}}
\newcommand{\tcr}{\textcolor{red}}
\newcommand{\tcb}{\textcolor{blue}}

\newcommand{\tcg}{\textcolor{ForestGreen}}

\global\parskip 6pt

\renewcommand{\vec}[1]{\boldsymbol{#1}}

\newcommand{\qed}{\nobreak \ifvmode \relax \else\ifdim\lastskip<1.5em 
\hskip-\lastskip\hskip1.5em plus0em minus0.5em \fi \nobreak\vrule height0.75em 
width0.5em depth0.25em\fi}

\newcommand{\be}{\begin{eqnarray}}
\newcommand{\ee}{\end{eqnarray}}

\def\>{\rangle}
\def\<{\langle}

\def\tr{\hbox{Tr}}

\newcommand{\executeiffilenewer}[3]{%
\ifnum\pdfstrcmp{\pdffilemoddate{#1}}%
{\pdffilemoddate{#2}}>0%
{\immediate\write18{#3}}\fi%
}

\graphicspath{{pics/}}

\usepackage{hyperref}

\begin{document}

\begin{titlepage}
\hfill{MPP-2015-248}
\\
\text{\ }\hfill{FPAUO-15/15}
\vspace*{.5cm}
\begin{center}
{\Large{\bf Entanglement Entropy in a Holographic Kondo Model}} \\[.5ex]
\vspace{1cm}
Johanna Erdmenger$^{a,}$\footnote{jke@mpp.mpg.de},
Mario Flory$^{a,}$\footnote{mflory@mpp.mpg.de},
Carlos Hoyos$^{b,}$\footnote{hoyoscarlos@uniovi.es},\\
Max-Niklas Newrzella$^{a,}$\footnote{maxnew@mpp.mpg.de} and
Jackson~M.~S.~Wu$^{c,d,}$\footnote{jknw350@yahoo.com}
\vspace{0.5cm}\\
\em{$^a$Max-Planck-Institut f\"ur Physik (Werner-Heisenberg-Institut),\\
F\"ohringer Ring 6, D-80805 Munich, 
Germany.
\\
$^{b}$Department of Physics, Universidad de Oviedo, 
\\
Avda.~Calvo Sotelo 18, 33007, Oviedo,
Spain.
\\
$^{c}$Department of Physics and Astronomy,
University of Alabama,
\\
Tuscaloosa, AL 35487, USA.
\\
$^{d}$National Center for Theoretical Sciences, Physics Division,
\\
No. 101, Section 2, Kuang Fu Road, Hsinchu, Taiwan 300, R.\,O.\,C.
}
\end{center}
\vspace*{1cm}
\begin{abstract}
We calculate entanglement and impurity entropies in a recent holographic model 
of a magnetic impurity interacting with a strongly coupled system. There is an 
RG 
flow to an IR fixed point where the impurity is screened, leading to a decrease 
in impurity degrees of freedom. This information loss corresponds to a volume 
decrease in our dual gravity model, which consists of a codimension one 
{hypersurface} embedded in a BTZ black hole background in three dimensions. 
There are matter fields defined on this {hypersurface} which are dual to Kondo 
field theory operators. In the large $N$ limit, the formation of the Kondo cloud 
corresponds to the condensation of a scalar field. The entropy is calculated 
according to the Ryu-Takayanagi prescription. This requires to determine the 
backreaction of the {hypersurface} on the BTZ geometry, which is achieved by 
solving the Israel junction conditions.  We find that the larger the scalar 
condensate gets, the more the volume of constant time slices in the bulk is 
reduced, shortening the bulk geodesics and reducing the impurity entropy. This 
provides a new non-trivial example of an RG flow satisfying the $g$-theorem. 
Moreover, we find explicit expressions for the impurity entropy which are in 
agreement with previous field theory results for free electrons. This 
demonstrates the universality of perturbing about an IR fixed point. 
\begin{center}
\textit{Dedicated to the memory of our colleague Peter Breitenlohner}
\end{center}
\end{abstract}
\vspace*{.25cm}
\end{titlepage}

\tableofcontents
\thispagestyle{empty}

\section{Introduction and Summary}
\label{sec::Intro}

The Kondo model, proposed by Jun Kondo in 
1964~\cite{Kondo01071964}, 
describes the interaction of a single magnetic impurity with the fermionic 
quasi-particles of a Landau Fermi Liquid (LFL). The Kondo Hamiltonian 
includes two terms: a kinetic term for the LFL quasi-particles, and a Kondo 
coupling between the LFL spin current and the impurity's spin.

The Kondo model was seminal in many respects. For example, the Kondo coupling 
constant approaches 
zero in the ultraviolet (UV), appears to diverge at a dynamically-generated 
scale and in the 
infrared (IR), conventionally expressed as a temperature, the 
Kondo temperature, 
$T_{\textrm{K}}$. The Kondo model thus provided an example of asymptotic freedom 
and 
dimensional transmutation pre-dating Quantum Chromodynamics (QCD) by almost ten 
years. Moreover, 
the Kondo model was seminal to Wilson's development of (numerical) 
Renormalization Group (RG) 
techniques~\cite{Wilson:1974mb,PhysRevB.21.1003,PhysRevB.21.1044}, 
integrability~\cite{PhysRevLett.45.379,Wiegmann:1980,RevModPhys.55.331,
doi:10.1080/00018738300101581,0022-3719-19-17-017,1994cond.mat..8101A,
ZinnJustin1998, 
PhysRevB.58.3814}, large-$N$ 
limits~\cite{PhysRevB.35.5072,RevModPhys.59.845,1997PhRvL..79.4665P,
1998PhRvB..58.3794P,2006cond.mat.12006C,2015arXiv150905769C}, Conformal Field 
Theory (CFT) 
techniques~\cite{1998PhRvB..58.3794P,Affleck:1990zd,Affleck:1990by,
Affleck:1990iv,Affleck:1991tk,
PhysRevB.48.7297,Affleck:1995ge}, and 
more~\cite{Hewson:1993,doi:10.1080/000187398243500}.

The ``Kondo problem'' is to find the eigenstates of the Kondo Hamiltonian for 
all values of 
temperature $T$, and from them to compute all observables. The Kondo problem is 
considered solved, 
via a combination of the techniques mentioned above. The solution reveals 
the ``Kondo effect'', 
that is, the screening of the impurity spin by the LFL when $T \lesssim 
T_{\textrm{K}}$. In 
particular, the LFL quasi-particles form a screening cloud around the impurity, 
the ``Kondo 
cloud'', which at $T=0$ has a characteristic size, $\xi_{\textrm{K}} = 
v/T_{\textrm{K}}$ where $v$ is the Fermi velocity. 
For a dilute concentration of magnetic impurities, the Kondo effect leads to a 
characteristic 
$-\ln(T/T_{\textrm{K}})$ contribution to the resistivity~\cite{Kondo01071964}.
At $T \approx T_{\text{K}}$, perturbation theory breaks down. 

Although the Kondo problem is considered solved, many questions about the 
original Kondo model 
actually remain very challenging, such as the  the dependence of the 
Entanglement Entropy (EE) on 
the size of a spatial 
sub-system\cite{1742-5468-2007-01-L01001,1751-8121-42-50-504009,
1742-5468-2007-08-P08003,
PhysRevB.84.041107},  the
effect of a quantum quench of the Kondo coupling \cite{2011arXiv1102.3982L},
or how to define and measure $\xi_{\textrm{K}}$ 
precisely~\cite{2009arXiv0911.2209A}. More generally, a major challenge is to 
solve the Kondo 
problem when the LFL is replaced by strongly-interacting degrees of freedom, 
including in 
particular, in one spatial dimension, a Luttinger liquid.

The Anti-de Sitter/CFT (AdS/CFT) 
correspondence~\cite{Maldacena:1997re,Gubser:1998bc,Witten:1998qj}, also known 
as holography, may be uniquely well-suited to address many of these 
open questions. AdS/CFT equates a weakly-coupled theory of gravity in 
$(d+1)$-dimensional AdS 
spacetime, $AdS_{d+1}$, with a strongly-coupled $d$-dimensional CFT ``living'' 
on the boundary of 
$AdS_{d+1}$. Typically the strongly-coupled CFT is a non-Abelian gauge theory in 
the 't Hooft 
large-$N$ limit~\cite{Aharony:1999ti,Aharony:2008ug}.

Various holographic impurity models have been proposed: see for example 
refs.~\cite{Kachru:2009xf,
Kachru:2010dk,Faraggi:2011bb,Jensen:2011su,Karaiskos:2011kf,Harrison:2011fs,
Benincasa:2011zu,
Benincasa:2012wu,Faraggi:2011ge,Itsios:2012ev,Erdmenger:2013dpa}. In all of 
these models, the 
global $SU(2)$ spin symmetry is replaced by an $SU(N)$, which is then gauged, 
that is, $SU(N)$ 
gauge fields were introduced, and possibly additional fields, such as 
supersymmetric partners of 
the gauge fields. The impurity is then described as an $SU(N)$ Wilson 
line~\cite{Maldacena:1998im,
Rey:1998ik,Camino:2001at,Yamaguchi:2006tq,Gomis:2006sb,Gomis:2006im}. Gauging 
$SU(N)$ introduces 
an additional coupling, the 't Hooft coupling. All of the models in 
refs.~\cite{Kachru:2009xf,
Kachru:2010dk,Faraggi:2011bb,Jensen:2011su,Karaiskos:2011kf,Harrison:2011fs,
Benincasa:2011zu,
Benincasa:2012wu,Faraggi:2011ge,Itsios:2012ev,Erdmenger:2013dpa} employed the 't 
Hooft large-$N$ 
limit and large 't Hooft coupling. In other words, all of the holographic models 
in 
refs.~\cite{Kachru:2009xf,Kachru:2010dk,Faraggi:2011bb,Jensen:2011su,
Karaiskos:2011kf,
Harrison:2011fs,Benincasa:2011zu,Benincasa:2012wu,Faraggi:2011ge,Itsios:2012ev,
Erdmenger:2013dpa} 
are strongly-interacting mean-field models.

However, most of the holographic models in 
\cite{Kachru:2009xf,Kachru:2010dk,Faraggi:2011bb,
Jensen:2011su,Karaiskos:2011kf,Harrison:2011fs,Benincasa:2011zu,Benincasa:2012wu
,Faraggi:2011ge,
Itsios:2012ev,Erdmenger:2013dpa} describe only fixed points. Indeed, only the 
holographic model 
of \cite{Erdmenger:2013dpa} describes an RG flow with a Kondo coupling that 
exhibites asymptotic freedom and the appearance of $T_{\textrm{K}}$. 
The new aspects of \cite{Erdmenger:2013dpa} include a holographic description
of the flow  from a UV to an IR fixed point triggered by a
`double-trace' marginally relevant operator and a gravity dual of the
screening mechanism. The 
essential difference between this holographic model and
the original Kondo model is that instead of the free electron gas,
there is an inherently strongly coupled system which interacts with
the magnetic impurity. Moreover, just as in the models  \cite{Kachru:2009xf,
Kachru:2010dk,Faraggi:2011bb,Jensen:2011su,Karaiskos:2011kf,Harrison:2011fs,
Benincasa:2011zu,
Benincasa:2012wu,Faraggi:2011ge,Itsios:2012ev,Erdmenger:2013dpa} the 
gauge/gravity duality approach
requires a large $N$ limit for the $SU(N)$ spin group in 
\cite{Erdmenger:2013dpa}.
Due to this limit, the characteristic logarithmic rise of the
resistivity at low temperatures is absent. Instead, a leading
irrelevant operator analysis shows that the resistivity scales with temperature 
as $T^\Delta$, $\Delta$ being a real number greater than one. This is
reminiscent of a Luttinger liquid behaviour, and an interesting task
for the future will be to determine the Luttinger parameter for the
holographic model proposed in \cite{Erdmenger:2013dpa}.

The large $N$ Kondo model was studied in field theory (with free
electrons and a global $SU(N)$ spin group) already some
time ago \cite{PhysRevB.58.3794,PhysRevLett.90.216403}. 
As was noted there, the model simplifies considerably by introducing slave 
fermions $\chi$ and writing the impurity spin operator $\vec{S}$ as $ S^a = 
\bar \chi T^a \chi$. 
$T^a$ is a 
generator of $SU(N)$ in the fundamental representation. 
$\vec{S}$ is in a totally antisymmetric representation, 
corresponding to a Young tableau with $Q$ boxes.
The chiral fermions are constrained by the fact that their charge 
density has to be equal to $Q$. The crucial property of the large $N$ Kondo 
model is that the screening corresponds to a condensation of the operator
${\cal O} = \psi^\dagger \chi$, where $\psi$ is an electron field and
$\chi$ the slave fermion. The condensate breaks the $U(1)\times U(1)$
symmetry of the electrons and slave fermions to the diagonal $U(1)$. 
In the large $N$ limit, long-range
fluctuations are suppressed, such that condensate formation is
possible also in two dimensions.
There is a phase transition at a 
critical temperature $T = T_c$ in the large $N$ Kondo model, whereas for finite 
$N$ the transition is a crossover. 

This condensation has a very natural analogue in gauge/gravity
duality, in the form of a holographic superconductor in AdS$_2$. 
Motivated by a top-down brane construction which realises the scalar
dual to the operator ${\cal O} = \psi^\dagger \chi$ as strings streching from 
D5 to D7-branes, in 
\cite{Erdmenger:2013dpa} a bottom-up holographic Kondo model was suggested, in 
which the electron current is dual to an AdS$_3$ Chern-Simons field $A_{\mu}$, 
in 
addition to the scalar field dual to ${\cal O}$ which lives on an AdS$_2$ 
subspace. 
Moreover, the charge density dual is given by  the temporal component
$a_t$ of a gauge field on AdS$_2$. Finite temperature is introduced by
considering a BTZ black hole background instead of AdS$_3$. In this model, the 
defect RG flow is generated by the gravity dual of the product operator ${\cal 
O} {\cal   O}^\dagger$, which is obtained in analogy to the holographic
double-trace deformation proposed in \cite{Witten:2001ua}. This RG flow
displays the required scale generation. Moreover, at low energies
${\cal O}$ condenses. This gives rise to a screening of the impurity,
as may be checked by calculating the flux of $a_t$ in the AdS$_2$ subspace. At 
the boundary, this flux coincides with $Q$. At the horizon, this flux decreases 
when lowering the temperature, such that there are less impurity degrees of 
freedom visible.

In this paper, in view of providing further comparison with field
theory results, we calculate the entanglement entropy for the model
described above. The application of the Ryu-Takayanagi (RT) approach 
\cite{Ryu:2006bv,Ryu:2006ef} for the
holographic entanglement entropy requires to include the effect of the
backreaction of the AdS$_2$ hypersurface on the AdS$_3$ geometry. 
For a simple holographic model which describes the backreaction of the
defect on the surrounding geometry, we follow 
\cite{Takayanagi:2011zk,Fujita:2011fp,Nozaki:2012qd,Erdmenger:2014xya} and 
consider a hypersurface with the required matter content, i.e.~a 
two-dimensional gauge field and a complex scalar. The geometry is then 
determined by the Israel junction conditions. As discussed in 
\cite{Erdmenger:2014xya}, these conditions are influenced by the 
energy conditions for the brane matter fields. 
In this paper, we will often refer to the hypersurface as ``brane'', 
as it is the remnant of the intersection of the D5 and D7-branes studied in 
the top-down model of \cite{Erdmenger:2013dpa}.
The geometry remains BTZ everywhere except at the locus of this
brane. The effect of the Israel junction conditions is to cut out
parts of the bulk volume, in dependence on the brane tension generated
by the brane matter fields.

We calculate the entanglement entropy in this geometry as follows: The
field theory entangling region is an interval of length $2 \ell$
symmetrically located on both sides of the impurity. Then, according
to the Ryu-Takayanagi prescription, the entanglement entropy is given
by the length of the bulk geodesic with endpoints coinciding with the
endpoints of the boundary interval. Via the Israel junction conditions
implementing the backreaction, the matter fields on the brane
affect the length of this geodesic, and thus the entanglement
entropy. 

Within field theory, the
 {\it impurity entropy} is defined by the difference 
\cite{1742-5468-2007-01-L01001,1751-8121-42-50-504009,1742-5468-2007-08-P08003,
PhysRevB.84.041107}
\begin{gather}
 S_{imp}(\ell)\equiv S(\ell)\big|_{\text{Impurity 
present}}-S(\ell)\big|_{\text{Impurity absent}} \, .
\label{SimpIntro}
\end{gather}
 This quantity was argued to provide a 
good description of the Kondo cloud profile
\cite{PhysRevB.84.041107}. We use the same definition for the impurity
entropy in our holographic computation. 
The subtraction guarantees a finite result in the holographic
approach.

A central
result of our holographic calculation is that as the 
condensate $\langle {\mO} \rangle = \langle \chi^\dagger
\psi \rangle$ increases, the spacetime volume shrinks. 
This results in 
shorter geodesics normal to the brane and hence in smaller impurity entropy. 
This is in agreement with field theory expectations, since in
the large $N$ limit, the formation of the $\langle \mO \rangle$
condensate corresponds to screening of the impurity. When the impurity
is screened, less impurity degrees of freedom are visible, which is indeed
the case in our holographic approach.
This physical property is encoded in the $g$-theorem 
\cite{Affleck:1991tk,Friedan:2003yc}. Indeed, our model adds one more example to 
the
explicit realizations of boundary RG flows manifestly satisfying the
$g$-theorem.  A field theory example for such a flow may be found in 
\cite{PhysRevA.74.050305}. For further examples within holography, see
\cite{Janik:2015oja,Takayanagi:2011zk,Fujita:2011fp}.
We also note that our holographic map between the decrease of entanglement 
entropy and the decrease of volume in the dual gravity theory may be related to 
recent discussions of quantum complexity in the holographic context 
\cite{Susskind:2014rva,Susskind:2014rvaAdd,Alishahiha:2015rta,Brown:2015bva,
MIyaji:2015mia}.

\begin{figure}[htb]
 \centering
 \def\svgwidth{0.35\columnwidth}
\executeiffilenewer{fittedbranesimple_latexed.svg}{fittedbranesimple_latexed.pdf}%
{inkscape -z -D --file=fittedbranesimple_latexed.svg %
--export-pdf=fittedbranesimple_latexed.pdf --export-latex}%
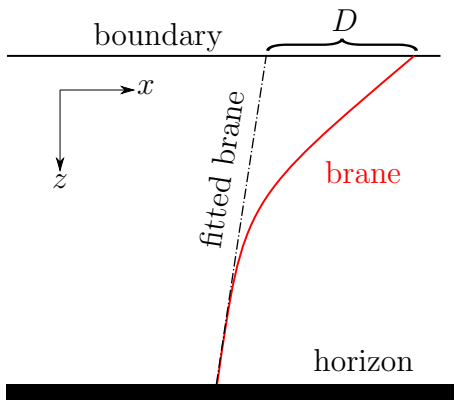%

\caption{Near horizon approximation of the brane via a constant tension 
 solution. See section \ref{sec::largeL} for details, and section 
 \ref{sec::Model} for the geometric construction of the backreaction of the 
 brane.}
\label{fig::fittedbranesimple}
\end{figure}

The agreement with field theory results may be made precise at least
in the large $\ell$ and small $T$  limit. The impurity entropy was calculated 
both using density matrix renormalisation group (DMRG) and CFT 
approaches 
\cite{1742-5468-2007-01-L01001,1742-5468-2007-08-P08003,
1751-8121-42-50-504009,PhysRevB.84.041107} for free electrons coupled
to a magnetic impurity. The $T=0$ result is 
\cite{1742-5468-2007-01-L01001,1751-8121-42-50-504009,1742-5468-2007-08-P08003,
PhysRevB.84.041107}
\begin{align}
S_{imp}=\frac{\pi c}{12}\frac{\xi_{\text{K}}}{\ell}\text{\ \ for\ \ 
}\ell\gg\xi_{\text{K}}.
 \label{AffleckT0}
\end{align}
Here $c$ is the central charge. In 
\cite{1742-5468-2007-08-P08003} this result was 
generalised 
to small but non-zero temperature,
\begin{align}
S_{imp}=\frac{\pi^2c\,\xi_{\text{K}} T}{6v}\coth\left(\frac{2\pi\ell 
T}{v}\right)=\frac{\pi^2c}{6}\frac{T}{T_{\text{K}}}\coth\left(2\pi\frac{\ell}{
\xi_{\text{K}}}\frac{T}{T_{\text{K}}} \right)
\text{\ \ for\ \ }T/T_{\text{K}}, \; \; \xi_{\text{K}}/\ell\ll1.
 \label{Affleck}
\end{align}
We find that a similar result is obtained in the holographic approach, where 
the 
field theory is strongly coupled. This is achieved in the following way: An RG 
flow is given by the defect brane which reaches from the boundary to the
horizon. Since the energy-momentum tensor
 originating from the defect matter fields
varies as function of the radial coordinate, the brane bends. 
It asymptotes to a constant tension brane in the UV near the boundary, i.e.~the 
energy-momentum tensor on the brane corresponds to constant tension.
Near the horizon, the energy-momentum tensor on the 
brane also asymptotes to constant tension form, however with a 
different tension than in the UV.
This corresponds to an IR fixed point. 
We consider a geometric linear perturbation about this fixed point
by  fitting a constant tension brane to the asymptotic behaviour of the brane 
at the horizon. 
Extrapolating this constant tension brane to the UV, it intersects
the AdS boundary at a distance $D$ from the impurity, see figure 
\ref{fig::fittedbranesimple}. 
The impurity entropy  \eqref{SimpIntro} then takes the form
\begin{align}
S_{imp}(\ell)=S_{\mathrm{BH}} (\ell +D) - S_{\mathrm{BH}} (\ell) \ , 
\end{align}
with $S_{\mathrm{BH}}$ the Ryu-Takayanagi entanglement entropy for the
BTZ black hole. Expanding to linear order in $D$,  
we recover the field theory result \eqref{Affleck} provided that we 
identify
\begin{align}
D \propto \xi_{\text{K}} \, ,
\end{align}
i.e. $D$ plays the role of the Kondo correlation length. This
remarkable agreement may be traced back to the fact that in both field
theory and holography, perturbation  about a CFT at an
IR fixed point is involved, which induces universal behaviour
independently of the precise form of the theory at the fixed point.

The structure of this paper is as follows: In section \ref{sec::Kondo} we 
summarise the most important features of our holographic Kondo model as 
presented in \cite{Erdmenger:2013dpa,Erdmenger:2014xya}. In particlular, we 
explain how the defect on the boundary is described by a thin brane in the 
bulk, and how the Israel junction conditions describe the backreaction of this 
brane on the geometry. In section \ref{sec::numericalresults} we present 
numerical results for the backreaction and entanglement entropy in our Kondo 
model. We then interpret these results, describing both the reduction of 
the impurity entropy and spacetime volume (interpreted as holographic 
complexity 
\cite{Susskind:2014rva,Susskind:2014rvaAdd,Alishahiha:2015rta,Brown:2015bva,
MIyaji:2015mia}) as 
signs of the screening of the impurity by the Kondo cloud. Furthermore, we
comment on the holographic $g$-theorem in section \ref{sec::g}, and compare our 
numerical and semi-analytical results to the field theory result \eqref{Affleck} 
in section \ref{sec::largeL}. As we will see, the 
null energy condition (NEC) plays a pivotal role throughout these discussions. 
Special attention is also given to the behaviour at $T=0$ in section 
\ref{sec::T0}. We then end 
in section \ref{sec::conc} with a conclusion and outlook. In appendix 
\ref{sec::JunctionCS} we comment on how to include the bulk Chern-Simons 
field in our calculations, and in appendix \ref{sec::NumericsApp} we give some 
details on the numerical methods employed to solve the equations of motion.

\section{Review: A Holographic Kondo Model}
\label{sec::Kondo}

\subsection{Action and geometric setup}
\label{sec::Model}

In this section we will briefly review the holographic model of the Kondo 
effect proposed in \cite{Erdmenger:2013dpa}, and the approach to including 
backreaction in this model outlined in \cite{Erdmenger:2014xya}. For
more details about this model, see those two papers. 
Results on the entanglement entropy obtained from this 
model will be discussed below in section 
\ref{sec::KondoEE}.
  
In \cite{Erdmenger:2013dpa}, a bottom-up holographic model was
motivated by a top-down string construction involving a background of
$N$ D3-branes as well as $k$ probe D7-branes and one probe D5-brane.
The probe D7-branes are embedded in such a way as to lead to chiral
fermions in $1+1$-dimensions 
\cite{Skenderis:2002vf,Harvey:2007ab,Harvey:2008zz,Buchbinder:2007ar}, 
which may be interpreted as electrons in
$3+1$ dimensions (for the interaction with the defect, the s-wave
approximation is used, as is standard in the CFT approach to the Kondo
effect). Note that the string construction makes them
inherently chiral. From the field theory point of view, these chiral fermions
correspond to a CFT with Kac-Moody algebra $SU(N)_k \times
SU(k)_N$.  Note that the $SU(N)_k$ symmetry is gauged in this
approach. The gauge anomaly is suppressed in the probe limit $k/N
\rightarrow 0$. The probe D5-brane introduces the $(0+1)$-dimensional
slave fermions $\chi$ as in \cite{Skenderis:2002vf,Camino:2001at,Gomis:2006sb}. 
In the dual gravity model, the D5-brane leads to the
instability inducing the condensation process.

Motivated by the above top-down construction, we now turn to a simpler
bottom-up model which allows to study the defect backreaction by
considering a thin brane. On the gravity side, this model contains
both $(2+1)$-dimensional bulk fields and $(1+1)$-dimensional defect
fields, where the defect spans the time and bulk radial directions. 
The starting point is a $(2+1)$-dimensional
Einstein-Hilbert contribution to the action, 
\begin{equation}
\mS_{g} = \frac{1}{2 \kappa_N^2} \, \int \! d^3 x \, \sqrt{-g} 
\left( R - 2 \Lambda \right)
\label{EHaction}
\end{equation}
with $R$ the Ricci scalar and the negative cosmological
constant $\Lambda=-1/L^2$, $L$ being the AdS radius. In the absence of 
the defect allows for the BTZ black hole solution
\cite{Banados:1992wn,Banados:1992gq}  with metric
\begin{align}
\totd s^2 = \frac{L^2}{z^2}\left( -f(z) \,\totd t^2 + \frac{\totd
    z^2}{f(z)} + \totd x^2 \right) \, ,
    \label{BTZlinelement}
\end{align}
where $L$ is the AdS radius and $f(z) = 1 - z^2/z_H^2$ with $z_H$ the position 
of the event horizon.
This provides  the finite temperature $T = 1/2 \pi z_H$ relevant for the Kondo 
effect.  Moreover, there is a
Chern-Simons (CS) field $A_{\mu}$ with action
\begin{equation}
 \mS_{A}=-\frac{\mN}{4\pi}\int \tr\left(A\wedge 
\totd A+\frac{2}{3}A\wedge A\wedge A\right)
\label{CSaction}
\end{equation}
defined throughout the (2+1)-dimensional bulk spacetime. 
Here, the
  normalization factor $\mN$ is 
proportional to $N$. As in the original model of \cite{Erdmenger:2013dpa},
$N$ corresponds to the rank of the gauged spin group $SU(N)$, which is
not directly visible in the bottom-up model considered.
Only the $SU(k)_N$ channel symmetry and an additional $U(1)$ charge
symmetry are explicitly realised by the Chern-Simons field. $N$ and
$k$ correspond to its level and rank.

We consider a single channel or
flavour, $k=1$, such that the Chern-Simons theory is $U(1)$. 
Moreover, there is a Yang-Mills (YM) field $a_m$, as 
well as a complex scalar field $\Phi$ localised in the worldvolume of a 
codimension one hypersurface that extends from the boundary to the black hole 
event horizon.
The scalar field is charged under both the CS and YM fields and carries 
opposite charges $\pm q$.
The action contribution for these fields reads
\begin{align}
&\mS_{a,\Phi}=- \mN \int \totd^2 x\, 
\sqrt{-\gamma}\,\left(\frac{1}{4}f_{mn}f^{mn} + 
\gamma^{mn}(\mD_m \Phi)^{\dagger} (\mD_n\Phi) +V(\Phi\Phi^{\dagger})\right)\, ,
\label{Kondobraneaction}
\\
&\mD_m\Phi \equiv \partial_m\Phi +i q A_m\Phi-i q a_m\Phi \, .
\label{covDer}
\end{align}
The precise form of the potential will be fixed later.
For the 
single-impurity Kondo model, the Yang-Mills field $a$ has gauge group $U(1)$. 
\footnote{A holographic model of the two-impurity Kondo effect 
which uses gauge group $U(2)$ was recently proposed in\cite{O'Bannon:2015gwa}.}
While Greek indices $\mu$ of bulk quantities run over 2+1 dimensions,
Latin indices $m$ of hypersurface quantities run over 1+1 dimensions,
i.e.~over time and the bulk radial direction. 
The induced metric on the hypersurface or `brane'  is denoted by
$\gamma_{mn}$. The precise form of the potential in \eqref{Kondobraneaction} will be fixed later.
Details on the projection of the CS field 
to the brane are given in 
appendix \ref{sec::JunctionCS}.  As 
shown in \cite{Erdmenger:2014xya}, the energy-momentum tensor on the brane 
worldvolume takes the form
\begin{align}
S_{ij}
&=\frac{\mN}{2}\gamma_{ij}f^{mn}f_{mn}
+2\mN\left[\left(\mD_{(i}\Phi\right)^{\dagger}\mD_{j)}\Phi
-\frac{1}{2}\gamma_{ij} 
\left(|\mD\Phi|^2+V\left(\Phi^\dagger\Phi\right)\right) 
\right ],
 \label{Sij}
\end{align}
where we used the antisymmetry of $f_{mn}$ to show 
$\gamma^{mn}f_{mi}f_{nj}=\frac{1}{2}\gamma_{ij}f^{mn}f_{mn}$ in our setup.

\begin{figure}[htb]
 \centering
 \def\svgwidth{1\columnwidth}
\executeiffilenewer{Kondo3_latexed.svg}{Kondo3_latexed.pdf}%
{inkscape -z -D --file=Kondo3_latexed.svg %
--export-pdf=Kondo3_latexed.pdf --export-latex}%
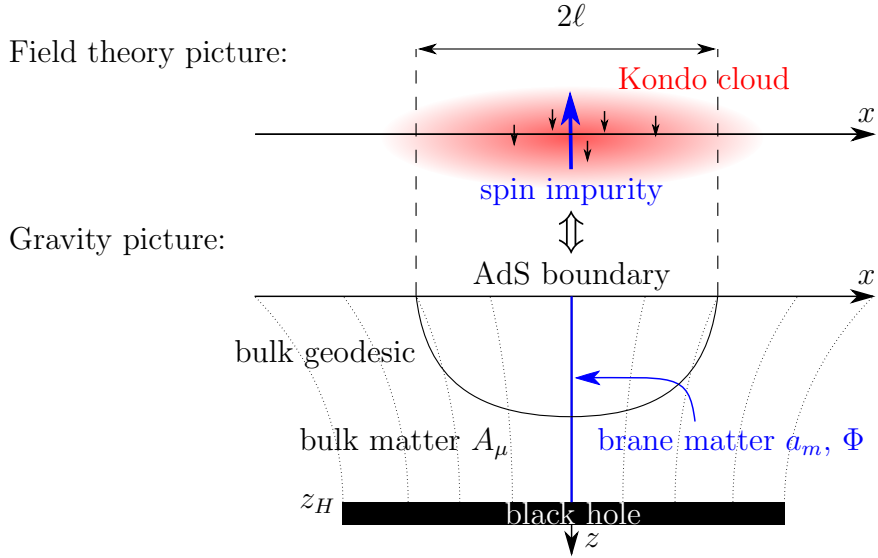%

\caption{Bulk setup of the holographic Kondo model of \cite{Erdmenger:2013dpa}. 
 The localised spin impurity on the field theory side is described by a 
 codimension one hypersurface (called `brane') in the bulk. In this figure, 
 the brane is shown to be trivially embedded, as appropriate for the case 
 without backreaction. We show the boundary interval of length $2\ell$ for 
 which we calculate the entanglement entropy. 
 The corresponding bulk geodesic used in the Ryu-Takayanagi prescription 
 streches into the bulk and crosses the impurity hypersurface.}
\label{fig::Kondo}
\end{figure} 

In \cite{Erdmenger:2013dpa}, where backreaction on 
the geometry was neglected, the embedding of the brane into the background 
spacetime is given by the function $x(z)\equiv0$, which we call the 
\textit{trivial embedding}. In order to calculate the entanglement entropy in this 
holographic model following the Ryu-Takayanagi (RT) approach 
\cite{Ryu:2006bv,Ryu:2006ef}, we have to include the backreaction of
the brane matter fields on the 
geometry. This is necessary in order to study the effect of the brane
matter fields on the bulk minimal surface over the entangling
region. In the low-dimensional case considered here, the minimal
surface entering the RT formula is just a geodesic.
Since the Chern-Simons bulk action is topological, the only
contribution to the energy-momentum tensor comes from the brane matter
fields $a_m$ and $\Phi$.

\begin{figure}[htb]
 \centering
 \def\svgwidth{0.7\columnwidth}
\executeiffilenewer{cutnpaste_latexed.svg}{cutnpaste_latexed.pdf}%
{inkscape -z -D --file=cutnpaste_latexed.svg %
--export-pdf=cutnpaste_latexed.pdf --export-latex}%
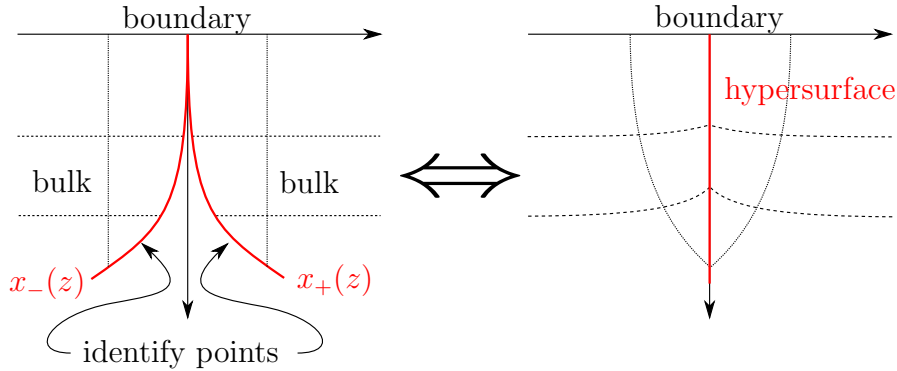%

\caption{Construction underlying the Israel junction conditions. Left: 
Construction using standard coordinates as in \eqref{BTZlinelement}. This is 
the approach folowed in this paper. Right: It is theoretically possible (albeit 
too complicated in general) to construct Gaussian normal coordinates around the 
brane.}
\label{fig::cutnpaste}
\end{figure}

The central point of our construction is that the effect of the brane onto the 
geometry is taken into account via the {Israel junction conditions} 
\cite{Israel:1966rt}. The geometric setup underlying these junction 
conditions is depicted in figure \ref{fig::cutnpaste}. On the left we see the 
approach followed in this paper, where standard coordinates are used as in 
\eqref{BTZlinelement}. 
The embedding of the brane is then described by the \textit{two} functions 
$x_{\pm}(z)$\footnote{In this entire work we are only interested in static 
setups, hence we neglect a possible time dependence $x_{\pm}(z,t)$.}. $x_+$ 
describes the embedding of the brane with respect to the spacetime on the right 
side of the brane, while $x_-$ describes the embedding with respect to the 
left side of the spacetime.
Corresponding points on these two curves have then 
to be identified. Hence the part of the spacetime inbetween these two curves 
is effectively cut out, leading to a reduction of the (regularised) volume of 
the spacetime. On the right hand side of figure \ref{fig::cutnpaste}, we show 
an alternative (but equivalent) picture based on Gaussian normal coordinates. 
In these coordinates, the brane and a neighbourhood on both sides around it can 
be described by the same coordinate patch (in contrast to the setup depicted on 
the left side of figure \ref{fig::cutnpaste}). In any case, the spacetime 
outside the brane is a vacuum solution of Einsteins equations, specifically 
the BTZ metric \eqref{BTZlinelement} in our model.
To summarise, we have two BTZ spacetimes glued together at the hypersurface 
(the brane). Depending on the tension of the matter fields on the brane, a 
certain part of the ambient BTZ spacetimes is removed, leading to a reduction 
of the regularised volume.

In the following we  assume a symmetric embedding 
($x_+=-x_-$), such that the Israel junction conditions take the form
\begin{equation}
 K^{+}_{ij} - \gamma_{ij} K^+ = 
-\frac{\kappa_N^2}{2}\,S_{ij}
 \label{eq:leftoverIsrael}.
\end{equation}
Here $K^{+}_{ij}$ is the extrinsic curvature tensor calculated from $x_+(z)$, 
with the convention that the normal vector points from the $-$ side to the $+$ 
side. The embedding $(t,z) \hookrightarrow (t,z,x_+(t,z))$ induces a metric 
$\gamma$ on the brane with line element
\begin{equation}
\totd s_{(\gamma)}^2 =  \frac{L^2}{z^2} \left( -f(z) \,\totd t^2 + \frac{1+f(z) 
\,x_+'(z)^2}{f(z)}\,\totd z^2\right).
\label{eq:inducedmetric}
\end{equation}
The gravitational coupling constant $\kappa_N^2$ in \eqref{eq:leftoverIsrael}, 
which is related to the Newton constant $G_N$ by $\kappa_N^2 = 8 \pi G_N$,  
is of order $1/N^2$. In order to have non-trivial backreaction on the geometry, 
we need to rescale all the fields to be of order $\sqrt{N}$, such that 
effectively $\kappa^2_N\mN$ is of order one, with $\mN$ as in \eqref{CSaction}.
This rescaling has to be accompanied by a corresponding rescaling of the charge 
$q$ in \eqref{covDer}.

A detailed discussion of this construction, as well as further 
references, may be found in \cite{Erdmenger:2014xya}. Moreover,
related equations have appeared in the literature already before 
in holographic studies of boundary CFTs, although through 
a different geometrical construction. The initial proposal may be
found in 
\cite{Takayanagi:2011zk,Fujita:2011fp,Nozaki:2012qd}\footnote{A sign 
difference between \eqref{eq:leftoverIsrael} and the convention used in 
these papers comes from the differing choice of orientation of 
the normal vector of the brane, while a factor of $1/2$ in our convention comes 
from the fact that we assume spacetime to be present to both sides of the 
defect, see figure \ref{fig::cutnpaste}. See \cite{Erdmenger:2014xya} for more 
details.}.
Applications are discussed in 
\cite{Alishahiha:2011rg,Setare:2011ey,Kwon:2012tp,Fujita:2012fp,
Nakayama:2012ed,Melnikov:2012tb,Astaneh:2014fga,Magan:2014dwa}. 
It should also be noted that the Israel junction conditions 
were used also in the study of domain walls and brane worlds, see 
\cite{Randall:1999vf,Karch:2000ct,Karch:2000gx,Battye:2001pb,Kraus:1999it,
Bowcock:2000cq}. 

In \cite{Erdmenger:2014xya}, some of us derived  general results for models 
described by equation \eqref{eq:leftoverIsrael}, as well as some simple 
exact solutions for models where the matter content on the brane is simpler 
than in \eqref{Kondobraneaction}\footnote{Specifically, we obtained exact 
solutions for the cases where the brane matter content is given by a constant 
tension, a YM-field in the absence of charges, a perfect fluid with equation 
of state $p=a\cdot\rho$, $a\geq0$ and a free massless scalar field, with 
background spacetimes being either AdS or the non-rotating BTZ black hole.}. In 
section 8 of \cite{Erdmenger:2014xya}, we also presented some preliminary 
results on the backreaction in the holographic Kondo model considered
also here. There, we neglected the effect of the Chern-Simons 
field, which we assumed to contribute to the equations of motion 
with its own junction conditions similar to
\eqref{eq:leftoverIsrael}. Here we verify this assumption by 
explicitly deriving these junction conditions in appendix \ref{sec::JunctionCS} 
and by showing that the CS field effectively decouples from the dynamics of the 
brane. This decoupling was implicit in the probe model of
\cite{Erdmenger:2013dpa} through a convenient gauge 
choice. Moreover, in the present paper, the numerical method and
precision is
substantially improved as compared to \cite{Erdmenger:2014xya}. 
For detailed information on the numerics, see appendix
\ref{sec::NumericsApp} below. In the
following, we briefly recapitulate the most important details of the
backreacted holographic Kondo model.

\subsection{Equations of motion and boundary conditions}
\label{sec::EOMs}
Above we discussed the Israel junction conditions which are the equations 
of motions for the embedding scalar $x_+$. In the following, we will discuss 
the equations of motions for the matter fields $\Phi$ and $a$, as well as their 
asymptotic behaviour.
We fix the scalar potential in \eqref{Kondobraneaction} to be a mass-term only,
\begin{align}
V(\Phi\Phi^{\dagger})=\frac{1}{2}M^2\Phi\Phi^{\dagger},
\label{eq:massterm}
\end{align}
see section \ref{sec::T0} for a further discussion of the potential. We 
choose the radial gauge $a_z = 0$ and use the residual gauge freedom to set the 
compex phase, $\psi$, of the scalar field $\Phi\equiv \phi \,e^{i\psi}$ 
to zero at some 
$z$ in the bulk. The radial equation 
for the gauge field implies that the phase has to vanish everywhere, which must 
be consistent with the boundary conditions for the scalar field. 
Hence we pick real boundary conditions for the scalar field.
The equation for the $z$-component of the gauge field is then trivially solved.
According to the discussion in appendix \ref{sec::JunctionCS}, the Chern-Simons 
field $A$ decouples from the matter confined to the brane, i.e.~the equations 
of motion for $\phi,\ a_t$ and $x_+(z)$ can be solved 
independently. They read
\begin{align}
 \gamma^{ij} \mD_{i} \mD_{j} \phi - M^2 \phi 
 &= 0, \label{eq:EOMscalar}
 \\
 \partial_z\sqrt{-\gamma}f^{zt} + J^t 
 &=0, \label{eq:EOMgauge}
\\
 \mK_{ij} - \frac{\kappa_N^2}{2} S_{ij} 
 &=0, \label{eq:EOMemb}
\end{align}
where the conserved current is
$J^{\mu} = - 2 \sqrt{-\gamma} \gamma^{\mu\nu} a_{\nu}\phi^2$ and $ 
\mathcal{K}_{ij}\equiv-\left(K^+_{ij} - \gamma_{ij} 
K^+\right)$. For the 
static case, we showed in \cite{Erdmenger:2014xya} that by projecting the 
tensorial equation \eqref{eq:EOMemb} to its trace and non-trace parts we obtain 
two scalar equations
\begin{align}
  \mathcal{K}=\frac{\kappa_N^2}{2}S\ \text{    and    
}\ \mathcal{K}_{L/R}=\frac{\kappa_N^2}{2}S_{L/R}.
  \label{EOMagain}
\end{align}
The index $L/R$ signifies the non-trace part and denotes ``left/right''.
This 
is because while the tensor that projects onto the trace part is by definition 
the metric ($S=\gamma^{ij}S_{ij}$), the tensors that project onto 
the traceless parts are constructed from the two independent null vectors  of 
the $1+1$-dimensional worldvolume of the brane, one pointing to the ``left'' 
($l^i$), one pointing to the ``right'' ($r^i$), with 
$l_ir^i\equiv-1$ and $S_{L}\equiv l^il^jS_{ij}$, $S_{R}\equiv r^ir^jS_{ij}$. In 
the static case where there is no energy flux from one side to the other, we 
can choose a coordinate system such that $S_L=S_R=S_{L/R}$ is 
the only component apart from the trace $S$ needed to define 
$S_{ij}$.

The two equations in \eqref{EOMagain} are not independent, 
but related due to conservation of energy-momentum \cite{Erdmenger:2014xya}. 
In particular, from \eqref{Sij} we find
\begin{align}
 S=\frac{\mN}{2}f^{mn}f_{mn}-2 \mN V\left(\Phi^\dagger\Phi\right),\ \ 
S_{L/R}&=\frac{\mN}{2}\widehat{\gamma}^{ij}\left(\mD_{(i}\Phi\right)^{\dagger}
\mD_
{ j) }\Phi.
\label{SandSLR}
\end{align}
In our choice of coordinates (c.f.~\eqref{eq:inducedmetric}), the positive 
definite tensor 
$\widehat{\gamma}^{ij}$ coincides with the Euclidean form of the metric 
$\gamma^{ij}$, see \cite{Erdmenger:2014xya}.
In order to ensure regularity at the event horizon we have to require
\begin{equation}
 \phi'(z_H) = -\frac{L^2\,M^2 \,\phi(z_H)}{2 z_H},\ \ \ 
 a_t(z_H)=0,\ \ \ 
x_+'(z_H) = \kappa_N^2 \mN\,\frac{2L^4 M^2  \,\phi(z_H)^2-z_H^{\,4} 
\,a_t'(z_H)^2}{4 
L^3}.
\label{horizonconsistency}
\end{equation}
The condition on the gauge field in the equation above 
comes from the fact that the Killing vector 
field $\partial_t$ has vanishing norm at the horizon. The other two conditions 
come from requiring that the most divergent expansion coefficients of the EOMs 
at the horizon should vanish.
Additionally, we need to impose  certain conditions at the boundary. The most 
trivial one is $x_{\pm}(0)=0$ for the embedding scalar. We can impose this
without loss of generality due to the translation invariance of the BTZ metric 
in the direction of $x$.

Treating the scalar as a probe with respect to the YM field, we find the 
leading behaviour
\begin{equation}
  a_t \sim Q/z + \mu + \ldots
  \label{eq:expansionAfirstorder}
\end{equation}
with $Q = - \mC L^2 \cosh(s/L)$ and $\mu=\mu_c = \mC L^2 
\cosh^2(s/L)/z_H$, where the length $s$ parametrises the embedding near the 
conformal boundary at $z = 0$, and $\mC$ is the electric flux of the YM field 
at the conformal boundary.
The length $s$ and its relationship with $\mC$ will be explained in more detail 
below in \ref{sec::consttension}, in particular 
in \eqref{backgroundSolutionGeodesicLength}.
In the top-down construction in \cite{Erdmenger:2013dpa}, it was shown that the 
electric flux $\mC$ coincides with the number of boxes in the Young tableau of 
the totally antisymmetric representation of the impurity spin $S^a = \bar \chi 
T^a \chi$.
$\mu$ is identified with the chemical potential for the $U(1)$ charge. 
For $\mu/T$ larger than a critical value, the scalar condenses.

For the scalar we find two asymptotic solutions near the boundary. As in 
\cite{Erdmenger:2013dpa,Erdmenger:2014xya}, we fix the mass $M$ in exactly 
such a way that the Breitenlohner-Freedman bound \cite{Breitenlohner:1982jf} is 
saturated in order to have the correct operator dimensions needed for a 
description of the Kondo effect. 
This yields
\begin{align}
 M^2 = \frac{4 Q^2 \cosh ^2\left(s/L\right)-1}{4 L^2 \,\cosh^2 
\left(s/L\right)} = \left(\frac{Q}{L}\right)^2 - (4 L^2 \,\cosh^2 
\left(s/L\right))^{-1}.
\label{BFbound}
\end{align}
This choice ensures that the scalar operator is marginally
relevant, such that there is asymptotic freedom.
The asymptotic behaviour of the scalar field hence becomes
\begin{equation}
 \phi(z) \sim \alpha \sqrt{z} \log(\Lambda\,z) + \beta \sqrt{z} + \ldots
\end{equation}
where we introduced an arbitrary energy scale $\Lambda$ to define the 
logarithm. Changing $\Lambda \rightarrow \Lambda'$ induces a renormalisation of 
the dimensionless coupling $\kappa = \alpha/\beta$\,  given by
\begin{equation}
\kappa(\Lambda') = 
\frac{\kappa(\Lambda)}{1+\kappa(\Lambda) \log(\Lambda/\Lambda')}\, .
\label{eq:RGcoupling}
\end{equation}
Next, we exploit two scaling symmetries of the EOMs to set $L=1$ and 
$z_H=1$. Coordinates and fields become dimensionless quantities which we denote 
by tilded variables\, 
\begin{align}
\tilde{z} = z/z_H,\quad 
\tilde{x} = x/z_H,\quad
\tilde{t}=t/z_H,\quad
\tilde{\phi} =\phi,\quad
\tilde{a}_t = a_t \,z_H,\quad
\tilde{x}_+ = x_+/z_H,\quad
\text{etc.}
\label{tilde}
\end{align}
In this new coordinate system we also have
$$
\tilde{a}_t \sim Q \frac{z_H}{z} + \mu\, z_H + \ldots
$$
and we define $\mu_T := \mu\, z_H = \mu / 2 \pi T$ which effectively labels the 
solutions for different temperatures.
Now
\begin{align*}
 \tilde{\phi}(\tilde{z}) = \phi(z) 
 \sim&\, \alpha \sqrt{z} \log(\Lambda z) + \beta \sqrt{z} +... \\
 =&\,\alpha\sqrt{z_H} \sqrt{\tilde{z}} \log(\tilde{z})
   + \sqrt{\tilde{z}}\sqrt{z_H}(\beta+\alpha\log(\Lambda z_H))\\
 =:&\,\alpha_T \sqrt{\tilde{z}}\log(\tilde{z})+\beta_T \sqrt{\tilde{z}}+\ldots
\end{align*}
We define $\kappa_T = \alpha_T/\beta_T$ and find the relationship
\begin{equation}
\kappa_T = \frac{\kappa}{1+\kappa\log(\Lambda z_H)} .
\label{kappaT}
\end{equation}
Using this relationship, we can define the Kondo temperature in the holographic 
model by demanding that \eqref{kappaT} diverges at this temperature for a given 
choice of $\kappa$ at some energy scale. It reads
\begin{equation}
T_{\text{K}} := \frac{\Lambda}{2 \pi} \,e^{1/\kappa(\Lambda)}
\label{KondoTempHolo}
\end{equation}
and hence
\begin{equation}
T/T_{\text{K}} = \exp(-1/\kappa_T)
\label{eq:ToverTK_vs_kappaT}
\end{equation} 
in the rescaled system. 
Note that $\kappa_T$ and the definition of $T_{\text{K}}$ are invariant 
under RG translations $\Lambda \rightarrow \Lambda'$, hence physical quantities.
Once the coupling constant $\kappa$ is chosen at some energy scale $\Lambda$, a 
Kondo temperature can be defined for the model.
That $\kappa$ indeed corresponds to the Kondo coupling constant on the 
field theory side is shown by holographic renormalisation in 
\cite{O'Bannon:2015gwa}. In this paper, we focus on the backreaction of the 
energy-momentum tensor on the brane. For this purpose, we do not need to 
renormalise our model since we do not compute any $n$-point functions or free 
energies.

To compute solutions for different ratios $\mu/T$, we increase $\mu_T$ and find 
that for $\mu_T > \mu_{T_c}$ the scalar field and thus $\alpha_T$ and $\beta_T$ 
become non-trivial and depend on $\mu_T$.
On the field theory side this corresponds to the condensation of the 
scalar operator $\langle \mO \rangle \neq 0$ below a critical temperature 
$T_c$.
In the limit $\mu_T \rightarrow \mu_{T_c}$ the value of $\kappa_T$ 
converges to $\kappa_{T_c}$ and we can use \eqref{eq:ToverTK_vs_kappaT} to 
define $T_c$ by 
\begin{equation}
T_c = T_{\text{K}} \, \exp \left(- 1/\kappa_{T_c}\right).
\label{eq:Tc}
\end{equation}
Numerically we find $\kappa_{T_c} \approx 8.16$ and hence $T_c 
\approx 0.885 \, 
T_{\text{K}}$, which is close to the values found in \cite{Erdmenger:2013dpa}.
To summarise, there is a phase transition at a temperature $T_c$ which is 
approximately the Kondo temperature $T_K$.
In this paper, we will not elaborate on this phase transition, which was 
described in more detail in \cite{Erdmenger:2013dpa}, but we find the same 
qualitative features.

\subsection{The normal phase}
\label{sec::consttension}

As explained in the introduction, in the large $N$ model considered, the 
screened phase is characterised by a condensate of the operator ${\cal
  O} = \psi^\dagger \chi$ dual
to the  scalar field $\Phi$. 
We refer to the screened phase with $\langle \mO \rangle \neq 0$ as the 
condensed phase, whereas we have $\langle \mO \rangle = 0$ in the normal 
phase. The UV  fixed point is
characterised by $\Phi=0, a_m\neq0$. As discussed in 
\cite{Erdmenger:2014xya}, at the UV fixed point the energy-momentum tensor 
corresponds to that of a constant tension brane, with  
$S_{ij}=-\frac{\mN}{2}\gamma_{ij}\mC^2$ where the 
constant\footnote{For an energy-momentum tensor of the form given here, the 
constancy of what is a priori the \textit{function} $\mC^2$ follows from the 
conservation of energy-momentum, $\nabla_i S^{ij}=0$.} is given by 
$\mC^2=-\frac{1}{2}f^{mn}f_{mn}$ and $\mC$ is the electric flux of the YM 
field as described above. Such constant tension solutions have been 
studied before in 
thin-brane AdS/BCFT models 
\cite{Azeyanagi:2007qj,Takayanagi:2011zk,Fujita:2011fp,Nozaki:2012qd,Erdmenger:2014xya}, 
and it is known that these solutions can be constructed by a
\textit{geodesic normal flow}, in an AdS or BTZ black hole 
background spacetime. This construction is obtained as
follows. 
The starting point is the the trivial embedding $x_+(z) = 0$ at the boundary. 
As $K_{\mu\nu}=0$ this embedding is totally geodesic, i.e.~geodesics in the 
worldvolume of the brane are also geodesics with respect to the ambient 
spacetime.
From every point on the brane, we construct a geodesic which is orthogonal to 
the brane at this point, which for our metric \eqref{BTZlinelement} can be 
done analytically.
We follow each of these geodesics for a certain arclength $s>0$ and define a 
family of embeddings $X_s$ for each value of $s$.
As explained in \cite{Erdmenger:2014xya}, each value of $s$ solves the Israel 
junction conditions for a specific constant tension on the brane. The gauge 
field on the brane is an explicit realisation of a constant tension solution if 
the scalar field vanishes.
By solving the Israel junction conditions \eqref{eq:leftoverIsrael}, we 
find a relationship between the parameter $\mC$ describing the gauge field 
solution and the parameter $s$ labelling the embedding, which is given by
\begin{align}
 \tanh\left(\frac{s}{L}\right)=\frac{L}{4}\kappa_N^2 \mN \mC^2 \, .
 \label{backgroundSolutionGeodesicLength}
\end{align}
This construction is very useful: The spacelike geodesics which enter  this normal 
flow construction coincide with those used for the Ryu-Takayanagi
entanglement entropy
prescription, if the entangling region is chosen to be an interval
placed symmetrically around the impurity. For a constant tension brane,
the results for the entanglement entropy hence take a 
very simple form. We will exploit this in section
\ref{sec::largeL} below. Similar results may be found in 
\cite{Azeyanagi:2007qj,Takayanagi:2011zk,Fujita:2011fp,Nozaki:2012qd,Erdmenger:2014xya}. 

In the normal phase where $\langle \mO \rangle = 0$, the Israel
junction conditions \eqref{eq:leftoverIsrael} as well as the 
gauge field equations of motion \eqref{eq:EOMgauge} may be solved analytically. 
For $\mC^2=-\frac{1}{2}f^{mn}f_{mn}$ and applying the radial gauge 
$a_z = 0$, we find
 \begin{align}
  a_t 	&= \frac{\mC L^2}{z_H} \cosh (s/L) 
     \left(\cosh(s/L)-\sqrt{(z_H/z)^2+\sinh^2 (s/L)}\right) \, ,
     \label{normalgauge}
     \\
  x_+(z)
  &=-z_H
  \,\arctanh\left(\frac{\sinh(s/L)}{\sqrt{(z_H/z)^2+\sinh^2(s/L)}}\right)
  \, .
   \label{backgroundSolutionGaugeAndEmbedding}
\end{align}
It should be noted that for 
$\mC^2>0$, the energy-momentum distribution on the brane described by 
\eqref{normalgauge},\eqref{backgroundSolutionGaugeAndEmbedding} 
satisfies the null and weak energy conditions (NEC and WEC) while violating the 
strong energy condition (SEC) which was for our $1+1$-dimensional brane matter 
defined in 
\cite{Erdmenger:2014xya} as
\begin{align}
(S_{ij}-S\gamma_{ij})m^{i}m^{j}\geq0\ \; \text{for any timelike vector $m^{i}$ 
tangent to the brane.}
\label{SEC}
\end{align}
While the NEC enters the 
proof of the holographic $g$-theorem 
\cite{Takayanagi:2011zk,Fujita:2011fp},   the violation of either WEC or 
SEC is required for our brane to reach the event horizon as in figure 
\ref{fig::Kondo} instead of bending back to the boundary. This was
shown in \cite{Erdmenger:2014xya}.

\section{Entanglement Entropy in the Holographic Model}
\label{sec::KondoEE}

\subsection{Numerical results}
\label{sec::numericalresults}
In this section we  present the results that we obtain by solving the 
equations of motion \eqref{eq:EOMscalar}-\eqref{eq:EOMemb}
numerically. Moreover, we present our numerical results for the
impurity entropy.

In the numerical calculations, the backreaction of the matter fields to the 
geometry is defined by fixing the ratio of the actions \eqref{EHaction}, 
\eqref{CSaction} and \eqref{Kondobraneaction} to $\kappa_N^2 \mN=1$.
The electric flux $\mC^2=-\frac{1}{2}f^{mn}f_{mn}|_{z=0}$ of the gauge field 
$a$ at the conformal boundary is set to 
$\mC = 1/2$ as in \cite{Erdmenger:2013dpa}, which fixes the leading behaviour 
of $Q$ in \eqref{eq:expansionAfirstorder}. Regularity at the horizon 
\eqref{horizonconsistency} fixes three 
more integration constants such that we are left with a one-parameter family of 
solutions, which we label by the value of the chemical potential $\mu_T = \mu 
/ 2 \pi T$. 
Starting at the critical value $\mu_T(T_c) = \mu_{T_c}$, see below 
\eqref{eq:expansionAfirstorder}, we choose a set of increasing values of 
$\mu_T$. For each of those, we integrate the 
EOMs numerically from the boundary to the horizon. While doing this,
we vary $\alpha_T$ and $\beta_T$ and until the regularity 
conditions \eqref{horizonconsistency} are satisfied. 
From this, we can compute the numerical embeddings at different ratios of 
$\mu / T$ or $T / T_c$, respectively.
More details about the 
numerics can be found in appendix \ref{sec::NumericsApp}. It should be 
pointed out that many of the results that we present in this and the following 
sections are qualitatively constrained by very general properties of the system 
such as by the geometry of the BTZ black hole or by energy conditions 
\cite{Erdmenger:2014xya}. Some of the results presented in section 
\ref{sec::largeL} will even be semi-analytical and applicable to more general 
models than our Kondo model. Hence, although we numerically fix $\kappa_N^2 
\mN=1$ we expect that our results represent the qualitative features of our 
model for a general value of this parameter. 

\begin{figure}[htb]
 \centering
 \def\svgwidth{0.8\columnwidth}
\executeiffilenewer{embeddingTurn_latexed.svg}{embeddingTurn_latexed.pdf}%
{inkscape -z -D --file=embeddingTurn_latexed.svg %
--export-pdf=embeddingTurn_latexed.pdf --export-latex}%
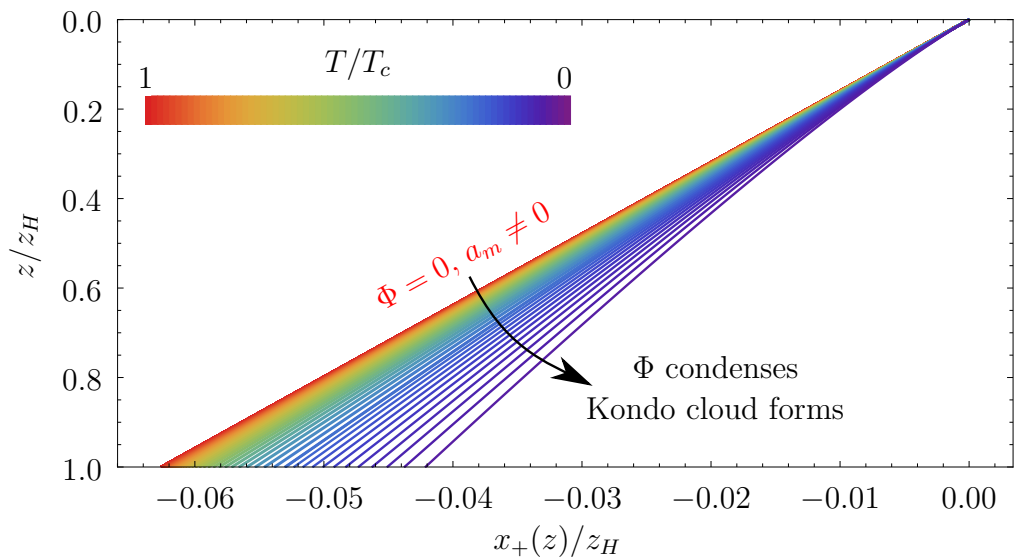%

\caption{Embedding profiles $x_+(z)$ for different values of $T/T_c$, 
which is adjusted numerically by changing $\mu_T$, see below 
\eqref{tilde}.
The curves start out at the left for $T=T_c$ and bend to the right as 
$T/T_c\rightarrow 0$. Compare to figure \ref{fig::cutnpaste} for the 
geometrical construction and the definition of $x_+(z)$. In particular, note 
that only the part of the spacetime to the right of the embedding curve is 
physical, while the left part is cut out and replaced by a mirror image of the 
right part.}
\label{fig::backreaction}
\end{figure}

First of all, the numerical solutions for $x_+(z)$ (more specifically 
$\tilde{x}_+(\tilde{z})$, see \eqref{tilde}) are shown in figure 
\ref{fig::backreaction}. For $T/T_c=1$ we are in the normal phase, 
for which $x_{+}$ is given by \eqref{backgroundSolutionGaugeAndEmbedding} which 
is the leftmost (red) curve in figure \ref{fig::backreaction}. As the 
temperature is lowered, the scalar $\phi$ condenses and hence adds a 
positive energy contribution (in the sense of the NEC) to the hypersurface 
energy-momentum tensor. As shown in \cite{Erdmenger:2014xya}, this leads to the 
embedding curves gradually bending more to the right, hence reducing the 
volume of the bulk spacetime. In fact, we found 
numerically that for all curves shown in figure \ref{fig::backreaction} the NEC 
and WEC are satisfied while the SEC (as defined in 
\eqref{SEC}) is violated for every $z$. This means that 
$x_{+}'(z)<0$ while $x_{+}''(z)\geq0$ for every 
$z$, and that the brane indeed reaches the event horizon instead of turning 
around and returning to the boundary \cite{Erdmenger:2014xya}.

We remind the reader that figure \ref{fig::backreaction} only shows 
$x_+(z)$ and our full geometry is constructed as shown in figure 
\ref{fig::cutnpaste}. This means that in figure \ref{fig::backreaction}, for any 
given temperature the spacetime to the left of the brane is cut out, and the 
remaining part to the right of the brane is glued along the brane to its mirror 
image. Hence as the curves bend to the right in figure 
\ref{fig::backreaction} for decreasing $T$,  the (renormalised) volume of the 
spacetime (or spacial slices thereof) is also decreasing. This behaviour was 
already commented upon for constant tension models in 
\cite{Azeyanagi:2007qj,Takayanagi:2011zk,Fujita:2011fp,Nozaki:2012qd,
Erdmenger:2014xya}. 

Let us now use these results to calculate the entanglement and
impurity entropies numerically.
Once the embedding functions $x_+(z)$ (and by symmetry assumption 
$x_-(z)=-x_+(z)$) are known, it is easy to 
holographically calculate 
entanglement entropies in the dual field theory using the RT prescription 
\cite{Ryu:2006bv,Ryu:2006ef}. This prescription states that for the $2+1$-dimensional 
background the entanglement entropy $S(A)$ of a boundary 
interval $A$ is given by
\begin{align}
 S(A)=\frac{c}{6}\frac{\mL_A}{L},
 \label{RT}
\end{align}
where $\mL_A$ is the length of a spacelike geodesic with its endpoints anchored 
to endpoints of the interval $A$, $c$ is the central 
charge of the boundary theory and we exploited the Brown-Henneaux formula $c = 
3 L /2 G_N$.

To calculate the entanglement entropy, it is advantageous that by 
construction (see figure \ref{fig::cutnpaste}) the bulk spacetime is still given 
by the BTZ metric except for the brane itself. For BTZ, solutions to the 
geodesic equations are known analytically. On the field theory side, we aim at 
calculating the effect of the Kondo cloud on the entanglement and impurity
entropies. For this purpose, we consider an interval of length $2\ell$ 
that is symmetrically centered around the defect. As the defect is located at 
$x=0$ on the boundary, we consider intervals $(-\ell,\ell)$ on the 
$x$-axis.\footnote{As we are interested in such symmetric intervals, the 
general refraction conditions for bulk geodesics at the brane 
reduce to the condition that the geodesics have to reach the brane 
at a right angle. See also 
\cite{Azeyanagi:2007qj,Takayanagi:2011zk,Fujita:2011fp,Nozaki:2012qd,Erdmenger:2014xya}.}  
This is shown in figure \ref{fig::Kondo}. 

In parallel to the field theory discussion given in the introduction
around \eqref{SimpIntro}, we define the impurity entropy also in the
holographic context by 
\begin{gather}
 S_{imp}(\ell)\equiv S(\ell)\big|_{\text{Impurity 
present}}-S(\ell)\big|_{\text{Impurity absent}} \, .
\label{Simpdef}
\end{gather}
For the case without defect, we simply
use the entanglement entropy in the BTZ background  accoring to the RT
prescription \cite{Ryu:2006bv}, 
\begin{align}
 S(\ell)\big|_{\text{Impurity absent}}=S_{BH}(\ell)=\frac{c}{3} 
\log\left(\frac{1}{\pi\epsilon T}\sinh\left(2\pi T 
\ell\right)\right).
\label{BTZresult}
\end{align}

\begin{figure}[htb]
 \centering
 \def\svgwidth{0.6\columnwidth}
\executeiffilenewer{Simp_latexed.svg}{Simp_latexed.pdf}%
{inkscape -z -D --file=Simp_latexed.svg %
--export-pdf=Simp_latexed.pdf --export-latex}%
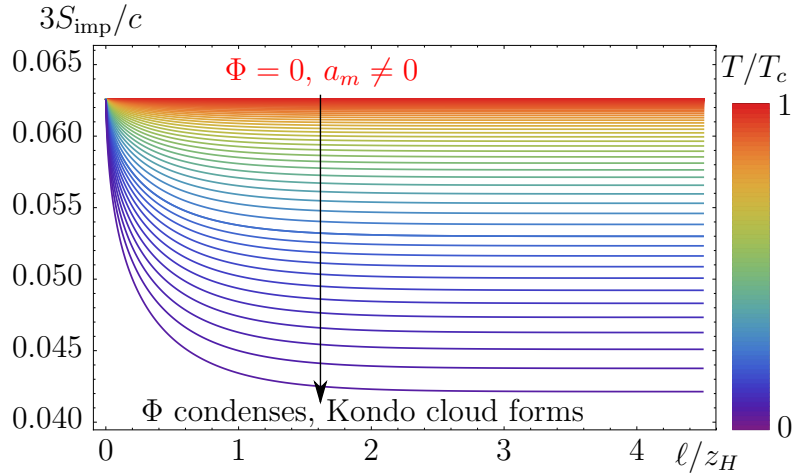%

\caption{Numerical results for the impurity entropy from the
  holographic model. For $T/T_c=1$ the normal phase is given be a constant 
tension solution \eqref{backgroundSolutionGaugeAndEmbedding} that leads to a 
constant $S_{imp}(\tilde{\ell})$ which is the uppermost curve in the figure. As 
the temperature is reduced and the scalar field condenses, due to the formation 
of the Kondo screening cloud the impurity entanglement for a given 
$\tilde{\ell}$ is reduced. }
\label{fig::Simp}
\end{figure}

The results for $S_{imp}(\ell)$ as defined in \eqref{Simpdef}  are 
shown in figure \ref{fig::Simp}. The interpretation of these numerical results 
is straightforward. We see that in the normal phase, i.e.~at $T/T_c=1$, we have 
$S_{imp}(\ell)=const.$ as for this case the embedding function 
$\tilde{x}_+(\tilde{z})$ corresponds to an effective constant tension solution 
\eqref{backgroundSolutionGaugeAndEmbedding}. Then as the 
temperature is lowered, the scalar field in the bulk becomes non-zero, which 
corresponds to the formation of the Kondo screening cloud. 
Due to the  screening, the entanglement is reduced for fixed 
$\tilde{\ell}$ as $T/T_c\rightarrow0$. From the bulk point of view, this is a 
consequence of following fact: Due to the NEC, the embedding curves 
$\tilde{x}_+(\tilde{z})$ increasingly bend to the right in figure 
\ref{fig::backreaction} as $T/T_c\rightarrow0$, and hence the bulk geodesics 
reaching the brane from the boundary point $\tilde{x}=+\tilde{\ell}$ have to 
travel a shorter distance, resulting in a reduced entanglement entropy. This can 
be interpreted as result of the Kondo cloud screening the impurity: The lower 
$T$ is for given $\tilde{\ell}$, the less impurity degrees of freedom are 
visible from outside the Kondo cloud.   Near the boundary the scalar field 
falls off such that for small $\tilde{z}$ coordinates, the embedding curves 
$\tilde{x}_+(\tilde{z})$ deviate only slightly from the constant tension 
solution of the normal phase, as is shown in figure \ref{fig::backreaction}. For 
this reason, for small $\tilde{\ell}$ the difference in entanglement entopy 
between condensed and normal phase goes 
to zero as displayed in figure~\ref{fig::Simp}. 

\begin{figure}[htb]
 \centering
 \def\svgwidth{0.7\columnwidth}
\executeiffilenewer{exp_latexed.svg}{exp_latexed.pdf}%
{inkscape -z -D --file=exp_latexed.svg %
--export-pdf=exp_latexed.pdf --export-latex}%
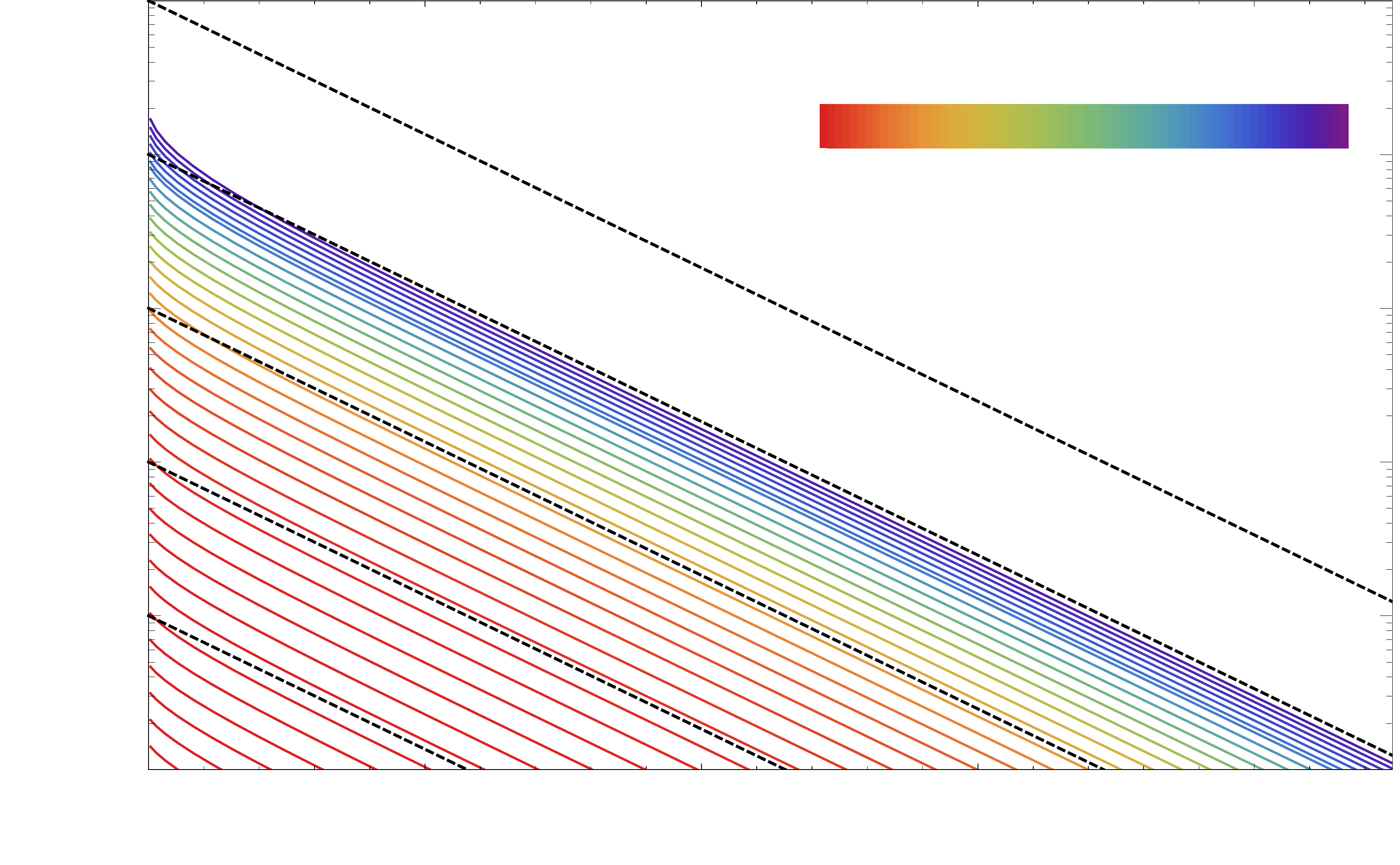%

\caption{Exponential falloff of $S_{imp}(\tilde{\ell})$ toward the limit value 
$S_{imp}(\tilde{\ell}\rightarrow\infty)$ for large $\tilde{\ell}$. The 
temperature decreases 
from the bottom to the top curve. The dashed (black) lines 
show an exponential falloff of the form $const.(T)\cdot e^{-2\tilde{\ell}}$ 
for 
comparison.}
\label{fig::backreactionEELog}
\end{figure}

In figure \ref{fig::backreactionEELog} we take the results shown in figure 
\ref{fig::Simp}, subtract the limit value 
$S_{imp}(\tilde{\ell}\rightarrow\infty)$ 
and present the resulting curves in a $\log$-plot. The linear behaviour 
for large $\tilde{\ell}$ shows that our curves for $S_{imp}(\tilde{\ell})$ 
display an exponential falloff $\sim const.(T)\cdot e^{-2\tilde{\ell}}$ 
towards the constant limiting value $S_{imp}(\tilde{\ell}\rightarrow\infty)$. 
We see that while the exponent of the falloff behaviour is the same for all our 
curves, the prefactor changes as a function of $T$. This will be investigated 
more thoroughly and compared to the field theory results \eqref{Affleck} in 
section \ref{sec::largeL}, yet the qualitative interpretation is clear: The 
larger the distance to the impurity is ($\tilde{\ell}\rightarrow\infty$), the 
less prominent the influence of the impurity. This is due to the screening by 
the Kondo cloud.

It is tempting to relate the decrease of the volume of the bulk spacetime 
mentioned above to recent proposals concerning holographic 
measures of complexity, see 
\cite{Susskind:2014rva,Susskind:2014rvaAdd,Alishahiha:2015rta,Brown:2015bva,
MIyaji:2015mia}. The 
original definition purported that holographically, a measure for the 
complexity $\mathfrak{C}$ of a quantum state can be defined as
\begin{align}
 \mathfrak{C}\propto \frac{\mV}{LG_N},
\end{align}
where $\mV$ is the volume of a codimension one spacelike hypersurface in the 
bulk geometry. In \cite{Brown:2015bva} this was generalised to the conjecture 
that $\mathfrak{C}$ can be obtained from the integral of the action over a 
certain bulk region. A thin brane construction with Israel junction conditions 
was employed in this context in \cite{MIyaji:2015mia}. We leave a detailed 
discussion of holographic complexity in our Kondo model for the 
future. Nevertheless, just as we will shortly interpret the decrease of 
entanglement entropy with the formation of the Kondo cloud as a sign of the 
screening of the impurity by that cloud, it is conceivable that the decrease of 
holographic complexity might be interpreted in the same way. In particular, our 
holographic map relates the decrease of entanglement to a reduction of the 
volume in the dual gravity theory.

\subsection{Holographic \texorpdfstring{$g$}{g}-theorem}
\label{sec::g}

Let us now investigate whether and how our holographic model obeys the 
$g$-theorem that is an analogue to the famous $c$, $a$ and $F$-theorems 
for boundary CFTs 
\cite{Affleck:1991tk,Yamaguchi:2002pa,Friedan:2003yc,Takayanagi:2011zk}. 
The $g$-theorem is certainly the most precise formulation of the idea that 
the Kondo cloud is a screening cloud: Along the RG flow from UV to IR, we 
expect the impurity entropy, i.e. the effective number of impurity degrees 
of freedom  to decrease.

\begin{figure}[htb]
 \centering
 \def\svgwidth{0.5\columnwidth}
\executeiffilenewer{lng_latexed.svg}{lng_latexed.pdf}%
{inkscape -z -D --file=lng_latexed.svg %
--export-pdf=lng_latexed.pdf --export-latex}%
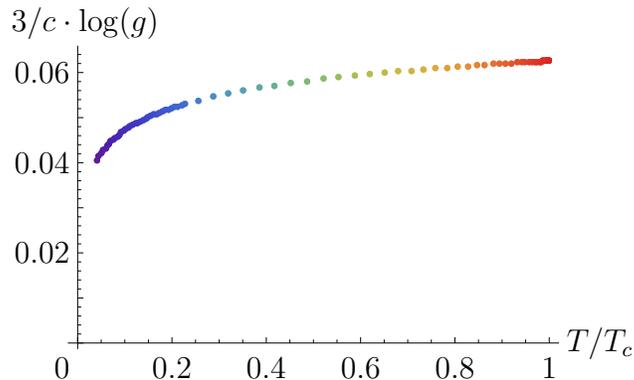%

\caption{Boundary entropy $\ln(g)$ as a function of temperature. }
\label{fig::lng}
\end{figure}

To do so, we first need to identify the boundary entropy from our numerical 
solutions, i.e. the contribution of the boundary (or defect) to the entire 
system. In the standard AdS/CFT case without defect, equation \eqref{BTZresult} 
defines the entanglement entropy for any interval of length 2$\ell$ on the 
boundary. As we take the limit $\ell\rightarrow\infty$ in which this interval 
spans all of the boundary, we enter a regime in which the 
entanglement entropy is linear in $\ell$ ($S_{BH}(\ell)\sim\ell$). In this 
regime, the entanglement entropy is not a good measure of quantum entanglement 
any more, but instead captures the thermodynamic entropy of the system which is 
of course expected to be an extensive quantity. From figure \ref{fig::Simp} we 
see that in this limit $\ell\rightarrow\infty$, the impurity entropy 
$S_{imp}(\ell)$ approaches a constant limiting value which can be interpreted as 
the contribution of the defect to the thermodynamic entropy of the
system. In fact, from the bulk geometry (see figure \ref{fig::backreaction}) we 
find that this additional entropy is just a result of the additional strip of 
the event horizon that we obtain because the brane is not located at the trivial 
embedding $x_+(z)=0$ any more. This trivial embedding corresponds to the 
absence of a defect. In particular,\footnote{\label{footnote123} Compared to 
\eqref{RT}, we take
$\mL_A=-2L\tilde{x}_+(z_H)$ since $\tilde{x}_+(z_H)$ is negative, and since
we obtain two additional strips of event horizon, one to the left and one to the 
right of the brane, see figures \ref{fig::Kondo} and \ref{fig::cutnpaste}. The 
correct amount of the additional horizon area is then $-2L\tilde{x}_+(z_H)$, 
not $-2Lx_+(z_H)$, due to the prefactor $\sim z^{-2}$ in the line element 
\eqref{BTZlinelement} evaluated at the horizon.}
\begin{align}
 S_{imp}(\ell\rightarrow\infty)=\frac{c}{3}\cdot (-\tilde{x}_+(z_H))
 \, ,
 \label{boundaryent}
\end{align}
where $\tilde x_+ \equiv x_+/ z_H$ as in \eqref{tilde}. 
Using the temperature as energy scale of the RG-flow (and identifying 
$\ln(g)\equiv  S_{imp}(\ell\rightarrow\infty)$), it was shown in 
\cite{Friedan:2003yc} that the $g$-theorem is equivalent to  the requirement 
that the impurity entropy increases with temperature (or decreases 
with inverse temperature),
\begin{align}
 T \cdot\frac{\partial S_{imp}(\ell\rightarrow\infty)}{\partial T}>0
 \label{lngT}
\end{align}
along the RG flow, which is obviously the case in our example as seen from figures 
\ref{fig::Simp} and \ref{fig::lng}. As explained in section 
\ref{sec::numericalresults}, this is a consequence 
of the null energy condition (NEC): The more the scalar field condenses, the more it 
contributes positive energy (in the sense of NEC) to the brane, and hence 
the more the brane bends to the right in the IR in figure
\ref{fig::backreaction} as compared 
to the normal phase. This means $|\tilde{x}_+(z_H)|$ decreases as the 
temperature decreases, and with \eqref{boundaryent} equation \eqref{lngT} 
follows straightforwardly.  

In fact, the NEC already played a central role in the holographic proofs of the 
$g$-theorem that were given in \cite{Takayanagi:2011zk,Yamaguchi:2002pa}. 
In \cite{Takayanagi:2011zk}, it was assumed that a thin brane described by 
\eqref{eq:leftoverIsrael} is embedded in a locally AdS spacetime as in the 
present paper, and the radial coordinate $z$ of the bulk spacetime was used as 
the scale of the RG flow. Let us use this interpretation here as well. Then, 
since the NEC is satisfied by the matter content of our holographic Kondo model, 
the $g$-theorem is guaranteed to hold in our case as well. As large $\ell$ on 
the boundary corresponds to large $z$ in the bulk, it is immediately obvious 
from figure \ref{fig::Simp} that $S_{imp}$ ($\equiv\ln(g)$ at conformal fixed 
points) decreases along the RG flow parametrised by $z$. 

\subsection{Large \texorpdfstring{$\ell$}{distance} approximation}
\label{sec::largeL}

As our results for the brane embeddings are numerical, so far (in
section \ref{sec::numericalresults}) we only presented numerical results for the 
impurity entropy as well. Yet, as we show now, in the limit of large $\ell$ it 
is possible to derive the form of the impurity entropy $S_{imp}$ due to very 
simple geometrical considerations.

\begin{figure}[htb]
 \centering
 \def\svgwidth{0.5\columnwidth}
\executeiffilenewer{fittedbraneNEW_latexed.svg}{fittedbraneNEW_latexed.pdf}%
{inkscape -z -D --file=fittedbraneNEW_latexed.svg %
--export-pdf=fittedbraneNEW_latexed.pdf --export-latex}%
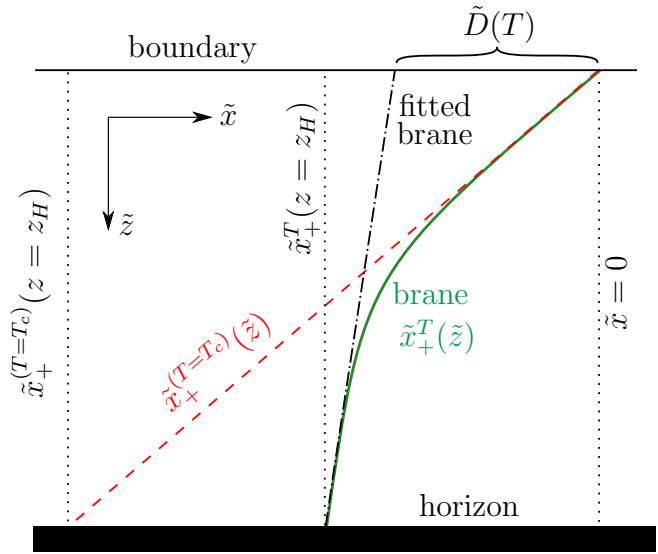%

\caption{Sketch of the construction used for obtaining the impurity
  entropy in a linear approximation about the IR at finite
  temperature. Near the horizon, the brane is approximated by a constant tension 
solution. For the normal phase with $\phi=0$ or  respectively at $T=T_c$, 
the brane is given by a constant tension solution (red dashed line). In 
contrast, for generic cases $T<T_c$ and $\phi\neq0$, due to violation of 
SEC and compliance with NEC the brane bends 
to the right until it reaches the event horizon at $z=z_H$ (solid green line). 
Fitting a constant tension solution to this generic brane near the horizon 
(black dot-dashed line), we define the quantity $D(T)$ as the distance between 
the point where this fitted brane meets the boundary and the point 
$\tilde{x}=0$ where 
the original brane started from the boundary. Although the constant 
tension solutions are given by an \textit{\arctanh} formula 
\eqref{backgroundSolutionGaugeAndEmbedding}, their curvature is not visible to 
the naked eye in the scale chosen for example in figure \ref{fig::backreaction} 
so that they appear as straight lines.}
\label{fig::fittedbrane}
\end{figure}

The key to this is that both near the boundary and near the horizon, our brane 
may be approximated by a constant tension solution, i.e. a solution to equation 
\eqref{eq:leftoverIsrael} with $S_{ij}\equiv\frac{S}{2}\gamma_{ij}$. The reasons 
for this are simple: First, near the boundary the scalar field falls off 
($\phi(z)\rightarrow0$), so that only the Yang-Mills field is left over, 
which describes a constant tension solution as explained in 
\cite{Erdmenger:2014xya} and in section \ref{sec::consttension}. In the case 
$\phi=0\Leftrightarrow T/T_c=1$, this approximation becomes exact throughout the 
bulk. Second, near the event horizon a constant tension solution can always be 
fitted to the brane, in particular as due to the consistency conditions 
\eqref{horizonconsistency} and the vanishing of some components of 
$\widehat{\gamma}^{ij}$ at the horizon we find from \eqref{SandSLR} that the 
non-trace part of the energy-momentum tensor behaves as $S_{L/R}\rightarrow0$ as 
$z\rightarrow z_H$. This constant tension brane then reaches the boundary at a 
shifted position $x=-D$ instead of $x=0$. This situation is shown in figure 
\ref{fig::fittedbrane}. 

As stated in section \ref{sec::consttension}, the constant tension solutions can 
be constructed from a geodesic normal flow, using the same geodesics that are 
also used in the Ryu-Takayanagi prescription to calculate entanglement entropy 
for symmetric intervals around the impurity. 
Hence, if the standard result for the entanglement entropy of an interval of 
length $2\ell$ in a BTZ background is \eqref{BTZresult}, then the result for a 
constant tension brane will read, with tension $\lambda$ embedded in the BTZ background 
such that $x_+(0)=0$,  
\begin{align}
 S_{EE}^{\lambda}(\ell)=S_{BH}(\ell)+C(\lambda).
 \label{backgroundSolutionEE}
\end{align}
The larger $\ell$ is chosen to be, the deeper the entangling curves penetrate 
into the bulk, and accordingly the closer to the horizon they encounter the 
brane. Hence, the larger $\ell$ becomes, the more accurate it is to replace our 
brane by a constant tension brane with some tension $\lambda'$ and
anchored on the boundary at $x_+(z)\rightarrow -D$ as $z\rightarrow0$, see 
figure \ref{fig::fittedbrane}. In 
this approximation, the entanglement entropy reads
\begin{align}
 S_{EE}^{\lambda',D}(\ell)=S_{BH}(\ell+D)+C(\lambda').
 \label{consttensionapprox}
\end{align}
From this, \eqref{BTZresult} and the definition \eqref{Simpdef} it is easy to 
find
\begin{align}
S_{imp}(\ell)&=C(\lambda')+\frac{c}{3}
\log\left(\frac{\sinh\left(2\pi T 
(\ell+D)\right)}{\sinh\left(2\pi T \ell\right)}\right)
\label{logsinh}
\\
&\rightarrow C(\lambda') +\frac{c}{3}
\left[2\pi T D+\left(1-e^{-4\pi T D}\right)e^{-4\pi T 
\ell}+\mO\left(e^{-8\pi T\ell}\right)\right]\text{\ \ for\ \ 
}\ell\rightarrow\infty
\label{TaylorExp}
\\
&\sim C(\lambda') +\frac{c}{3}
\left[
2\pi T D
+4\pi T D
e^{-4\pi T \ell}+\mO\left(e^{-8\pi T\ell}\right)\right]\text{  for 
$\ell\rightarrow\infty$ and 
small $D$.}
\label{cothseries}
\end{align}

Interestingly, if we assume $D$ to be small compared to both $\ell$ and $z_H$, 
we can simply approximate 
$S_{BH}(\ell+D)-S_{BH}(\ell)\sim D\cdot \partial_{\ell} S_{BH}(\ell)$ and 
hence
\begin{align}
S_{imp}(\ell)\approx C(\lambda') +\frac{2\pi c}{3} 
 T D(T) \coth\left(2\pi T \ell\right) \text{ for 
$\ell T\gg1$ and $DT\ll1$,}
 \label{Holofleck}
\end{align}
where we reinstated the potential temperature dependence that $D$ may have on 
its own.\footnote{A 
similar 
argument can be made for the $T=0$ case. There we have 
$S_{AdS}(\ell)=\frac{c}{3}\log(\frac{2\ell}{\epsilon})$ instead of 
\eqref{BTZresult} and $D\cdot \partial_{\ell}
S_{AdS}(\ell)=\frac{cD}{3\ell}$ which can be compared to the $T=0$ result 
\eqref{AffleckT0}.}
Using $-4\pi T \ell=-2\ell/z_H$, the formula \eqref{TaylorExp} explains the 
exponential behaviour of our numerical results 
displayed in figure \ref{fig::backreactionEELog}, while identifying the 
temperature dependent prefactor of the falloff with $1-e^{-4\pi T 
D}$. Note that in contrast to equations \eqref{cothseries} and 
\eqref{Holofleck}, equation \eqref{TaylorExp} does not require to assume 
that $D$ is small.

This holographic result may be compared directly to the field theory result
\eqref{Affleck} for  $T\rightarrow0,\  
\ell\rightarrow\infty$.  First of all, the two 
expressions obviously differ in the constant boundary entropy term 
$C(\lambda')$, which itself may have a temperature dependence. It may in 
principle vanish as $T$ goes to zero. Second, the arguments of 
the \textit{coth} agree exactly if we set $v=1$, i.e.~if we assume that 
the Fermi 
velocity is equal to the speed of light. This is correct as we are working with 
massless chiral fermions in our model, see \cite{Erdmenger:2013dpa}. Finally, let 
us compare the prefactors of the \textit{coth} in both \eqref{Affleck} and 
\eqref{Holofleck}. In the case of exact equality (and with $v=1$), these two 
equations imply for $T\approx0$
\begin{align}
 D(T)\approx\frac{\pi}{4}\xi_{\text{K}}\Leftrightarrow 
\tilde{D}(T)\approx \frac{\pi^2}{2}\xi_{\text{K}}T,
 \label{D}
\end{align}
where we have used $2\pi T D =D/z_H= \tilde{D}$ in accordance with 
\eqref{tilde}.
Of course, as we explained in the introduction, there are significant
differences between the expected field theory in our holographic model
and the field theory models used for deriving \eqref{Affleck}. Nevertheless, our 
results suggest that the simple geometric quantity $D(T)$, easily defined as 
shown in the sketch in figure \ref{fig::fittedbrane}, should approach a constant 
at low $T$. Equivalently, $\tilde{D}(T)$ which can be read off from figure 
\ref{fig::backreaction} should show linear behaviour with $T$ for low 
temperatures. Then, $D(T=0)$ may be viewed as a measure for the Kondo 
scale 
$\xi_{\text{K}}$ (up to a factor of order one). This is a very striking 
prediction.

Do our numerical results verify this claim \eqref{D}? In order to answer this 
question, we first have to discuss how much the temperature can rise before the 
behaviour \eqref{D} is lost. As it turns out, there is a very simple 
geometrical 
argument for this as we now explain. In fact, $\tilde{D}(T)$ is bounded from 
above. Thus for 
temperatures higher than $T=T^*$, the linear behaviour \eqref{D} is no longer 
possible.

\begin{figure}[htb]
 \centering
 \def\svgwidth{0.95\columnwidth}
\executeiffilenewer{LogLogDvsT_latexed.svg}{LogLogDvsT_latexed.pdf}%
{inkscape -z -D --file=LogLogDvsT_latexed.svg %
--export-pdf=LogLogDvsT_latexed.pdf --export-latex}%
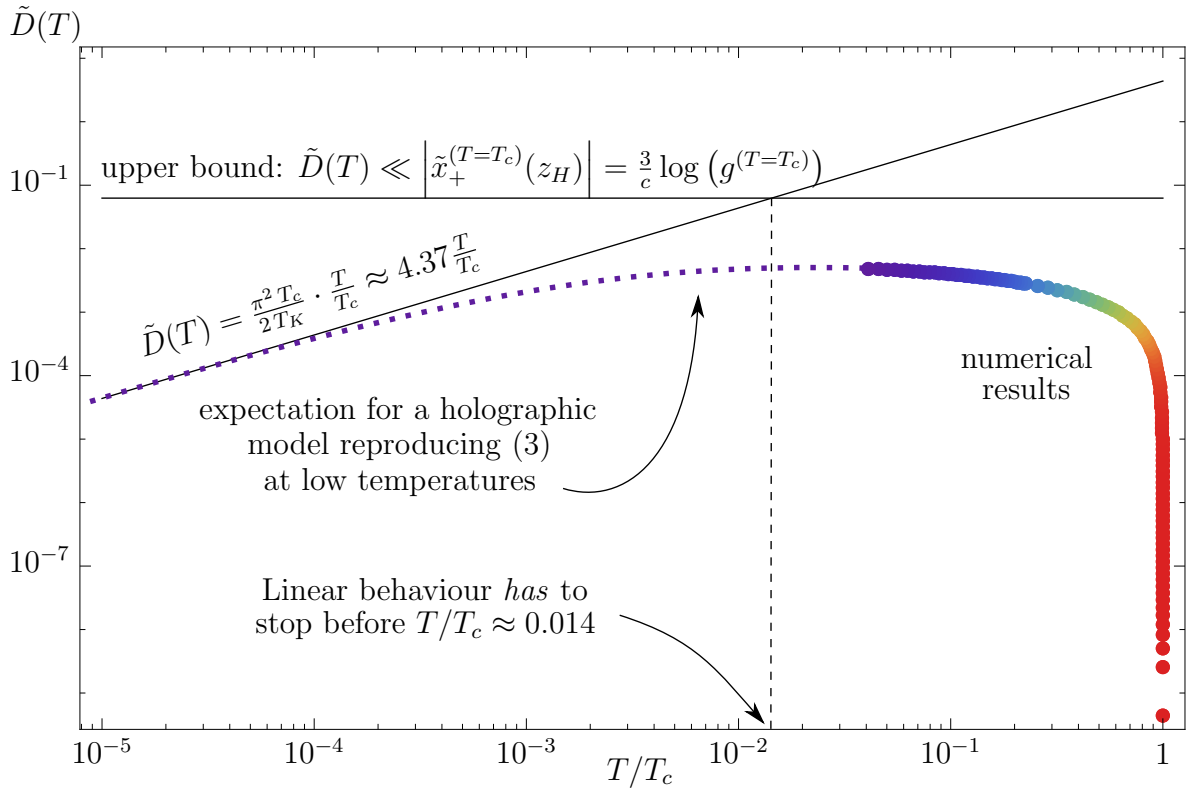%

\caption{Numerical results for $\tilde{D}(T) = 2\pi T D(T)$ as a function of 
$T/T_c$ (large points), together with the expected behaviour 
of a holographic model reproducing \eqref{Affleck} exactly at low $T$ 
(small points). As solid lines, we draw the upper bound on $\tilde{D}(T)$ given 
by \eqref{TDbound} as well as a model of the form 
$\tilde{D}(T)=\frac{\pi^2}{2}\frac{T}{T_{\text{K}}}$ which corresponds exactly 
to the 
field theory result \eqref{Affleck} as $T \rightarrow 0$.}
\label{fig::DvsT}
\end{figure}

The detailed argument for this is as follows. Due to the null energy condition 
(NEC), $\tilde{x}_+(z_H)$ of \eqref{tilde} can only increase along the boundary 
RG flow from the UV to 
the IR, i.e.~$|\tilde{x}_+(z_H)|$ decreases. 
Moreover, due to the violation of the strong energy condition (SEC), we will 
always have $\tilde{x}'_+(z_H)<0$. 
From this we derive that
\begin{align}
\tilde{D}(T) \leq |\tilde{x}^{(T)}_+(z_H)| \leq 
|\tilde{x}^{(T=T_c)}_+(z_H)|.
 \label{TDbound}
\end{align}
This equation is easily understood from figure \ref{fig::fittedbrane}: The 
interval $\tilde{D}$ is shorter than the $\tilde{x}$-interval defined by the 
point where the brane solution (in solid green) meets the horizon. In turn, 
this interval is smaller than the $\tilde{x}$-interval defined by the 
point where the UV constant tension brane (in red, dashed) meets the horizon. 
As explained in section \ref{sec::g}, the quantity 
$|\tilde{x}^{(T=T_c)}_+(z_H)|$ in \eqref{TDbound} has a very simple physical 
interpretation: It is half the length of the additional piece of event horizon 
that we obtain in the 
normal phase as compared to the BTZ black hole case without impurity (see 
footnote \ref{footnote123}). In particular, as 
explained in the previous subsection it is related to the impurity entropy in 
the normal phase by 
\begin{align}
 |\tilde{x}^{(T=T_c)}_+(z_H)|=\frac{3}{c}\ln \left(g^{(T=T_c)}\right).
 \label{lng}
\end{align}
From \eqref{backgroundSolutionGaugeAndEmbedding} and 
\eqref{backgroundSolutionGeodesicLength}, it follows that 
\begin{align}
|\tilde{x}^{(T=T_c)}_+(z_H)|= s /L = \arctanh \left(\frac{L 
\kappa_N^2  \mN \mC^2}{4}\right),
\end{align}
where $\mC$ here denotes the electric flux defined in 
section \ref{sec::Kondo}.
In our numerics, we set $\kappa_N^2  \mN =1$, $\mC=1/2$ and 
$L=1$, 
hence 
\begin{align}
 |\tilde{x}^{(T=T_c)}_+(z_H)|= \arctanh(1/16)\approx 0.0626,
\end{align}
a number featuring prominently in figures 
\ref{fig::backreaction}, \ref{fig::Simp} and 
\ref{fig::lng}. This 
is now very important: $\tilde{D}(T)$ may 
initially grow linearly with $T$ as in \eqref{D}, but if it is bounded from 
above by $\tilde{D}(T)\leq s/L$, this means that this linear behaviour has to 
stop (most 
likely long) before a temperature $T^*$ is reached 
where $\tilde{D}(T^*)=s/L$. The value of $T^*$ 
depends on the slope of the initial linear increase. If we assume for small 
$T/T_c$ that $\tilde{D}(T) \sim const. \frac{T}{T_c}$, our estimate is that the 
linear behaviour of $\tilde{D}(T)$ can under no circumstances be expected to be 
seen above 
\begin{align}
\frac{T}{T_c}\gtrsim \frac{T^*}{T_c}\approx 0.014,
\label{Tstar}
\end{align}
were we have used \eqref{TDbound} and by comparison with 
\eqref{D} we have set 
$const.=\pi^2 T_c/(2 T_{\text{K}})\approx 4.37$. As this 
is only a bound, the linear behaviour may already end at a temperature 
orders of magnitude lower than this, see also figure \ref{fig::DvsT}. 

To summarise this section, we have found that in certain limits (large 
$\ell$, small but finite $T$) we may qualitatively reproduce 
field theory results for the impurity entropy from simple geometrical
considerations. This applies in particular to the exponential falloff following 
from 
the field theory result \eqref{Affleck}. These arguments are analytical and only 
dependent on general geometric properties of our model (such as the BTZ 
background geometry that implies \eqref{BTZresult}). This indicates that our 
approach of approximating the curved brane of our holographic model by a 
constant tension brane near the event horizon (figure \ref{fig::fittedbrane}) 
corresponds to a linear perturbation about the IR fixed point. For this reason, 
we expect the results \eqref{TaylorExp} and, where the necessary approximations 
are valid, also \eqref{Holofleck} to be examples for universal behaviour of 
holographic models of AdS$_3$/BCFT$_2$ duality involving thin branes 
\cite{Takayanagi:2011zk,Fujita:2011fp,Nozaki:2012qd}, beyond the specific Kondo 
model studied here. Furthermore, it became apparent that while the \textit{coth} 
terms in \eqref{Affleck}, \eqref{Holofleck} agreed exactly in their respective 
arguments, we did not find even  qualitative agreement of the prefactors of the 
\textit{coth} term, as far as their dependence on temperature is concerned, see 
figure \ref{fig::DvsT}. In fact, we ended the section by showing that even if 
there \textit{were} an agreement of these prefactors, it could only be manifest 
at very low temperatures, lower than what we were able to study numerically. We 
believe that the argument we used to show this is interesting in its own right: 
While the equations \eqref{Affleck}, \eqref{Holofleck} were clearly derived 
under the assumption of low temperature ($T/T_c\ll1$), we did not
 specify what ``low'' means quantitatively 
in this context. However, in  \eqref{Tstar} we derived a bound on this \textit{low} temperature 
behaviour.
Curiously, this  bound was derived from equations \eqref{TDbound} and \eqref{lng},
which are dependent on the \textit{high} temperature ($T=T_c$) 
properties of our system. 

Based on these findings and especially on the results displayed in figure \ref{fig::DvsT}, it would 
certainly be interesting to study our system over several orders of 
magnitude of temperatue. However, not only has it turned out to be very demanding to 
extend our numerical results to low values of $T/T_c$ with adequate accuracy:
As we show in the next section, a physical behaviour at 
low temperatures can only be expected for more complicated forms of the potential 
$V(\Phi^\dagger\Phi)$ than the square term \eqref{eq:massterm} used so far.

\subsection{\texorpdfstring{$T=0$}{Zero temperature} behaviour}
\label{sec::T0}

As pointed out in the previous section, it is interesting for us how our system 
behaves as we approach $T=0$. This could be studied directly by embedding our 
brane into a \Poincare background instead of a BTZ metric as before, but 
unfortunately this is very demanding to do numerically. As we show in this 
section, we may extract some information about this case by using
analytical arguments. The zero temperature limit of holographic superconductors 
was studied for instance in \cite{Horowitz:2009ij}.

The main question of the $T=0$ case is: Will our system flow to a conformal 
fixed point? In our AdS/BCFT like ansatz, this would mean that at $T=0$ our 
system approaches a constant tension solution, at least deep in the bulk 
\cite{Takayanagi:2011zk,Fujita:2011fp,Nozaki:2012qd}. We would thus expect
\begin{align}
S_{ij}&\rightarrow const.\cdot\gamma_{ij} \text{\ \ as\ \ }z\rightarrow\infty
\\
\Rightarrow S_{L/R}&\rightarrow 0 \text{\ \ as\ \ }z\rightarrow\infty
\end{align}
for the energy-momentum tensor and components defined in \eqref{Sij} and 
\eqref{SandSLR}, if indeed we find a conformal IR fixed point. From the specific 
form of both $S_{L/R}$ and $\mK_{L/R}$ (see equation \eqref{SandSLR} and 
\cite{Erdmenger:2014xya}), this translates to the set of 
assumptions 
\begin{align}
  S_{L/R}\rightarrow0,\  
   \phi\rightarrow const.\neq0,\ 
   a_t\rightarrow 0,\ 
 x'(z)\rightarrow const.,\text{\ \ as\ \ }z\rightarrow\infty
  \label{goestozero}
\end{align}
where, as $\widehat{\gamma}^{ij}$ behaves as $\sim z^2$, we can additionally 
show that $x''(z)$, $\phi'$ and $a_t$ have to approach zero faster than 
$z^{-1}$. This 
has to be contrasted with the scalar equation of 
motion
\begin{align}
\partial_m\left(\sqrt{-\gamma} \gamma^{mn}\partial_n\phi\right)=\sqrt{-\gamma} 
\gamma^{mn}a_{n}a_{m}\phi+\frac{1}{2}\sqrt{-\gamma}\frac{\partial 
V}{\partial \phi},
\label{scalarEOMdetail}
\end{align}
where we used the same gauge choice as in 
\eqref{eq:EOMscalar}-\eqref{eq:EOMemb}.
Suppose that as $\phi\rightarrow const.\neq 0$ the term $\frac{\partial 
V}{\partial \phi}\neq0$. For our simple quadratic potential this is 
inevitable, but here we want to investigate the more general case. Due to the 
factor $\sqrt{-\gamma}$, the last term in \eqref{scalarEOMdetail} will then go 
to zero as $z^{-2}$ for $z\rightarrow\infty$. This means that 
\eqref{scalarEOMdetail} can only be satisfied if any other of the terms in this 
equation behaves exactly as $z^{-2}$ for large $z$, too. Unfortunately, from 
\eqref{goestozero} and our reasoning below this equation, we can actually 
deduce that all the other terms in \eqref{scalarEOMdetail} go to zero strictly 
faster than $z^{-2}$ for large $z$. This contradiction leads to two 
conclusions: First, for our simple model with potential $V\sim\phi^2$ the 
equations of motion seem to forbid the existence of an IR fixed point in our 
model. Second, for an IR fixed point to exist, it is necessary that for large 
$z$ $\phi\rightarrow\phi^{*}$ such that $\frac{\partial 
V}{\partial \phi}\big|_{\phi=\phi^{*}}=0$. This can easily be achieved by 
adding quartic or higher terms to the potential of our model. We then 
expect that our model with square potential is a good approximation to this 
improved model only when the maximal value of $\phi(z)$ is 
small compared to $\phi^*$. Differences between these two 
models would then only become apparent at low temperatures when the scalar 
field $\phi$ grows in the bulk so that the higher order terms in the potential 
become important. For now, we leave the study of a model with a more 
complicated potential for future research and continue to summarise the results 
obtained with our original simple holographic Kondo model as studied in 
\cite{Erdmenger:2013dpa,Erdmenger:2014xya}.   

\section{Conclusions and Outlook}
\label{sec::conc}

\subsection{Conclusion}
\label{sec::summ}

Let us recapitulate the main results of this paper.
In section \ref{sec::numericalresults} we have presented 
numerical results on the impurity entropy in the holographic model of the 
Kondo effect that was studied in 
\cite{Erdmenger:2013dpa,Erdmenger:2014xya} as 
summarised in section \ref{sec::Kondo}. We discussed how several qualitative 
features resulted straightforwardly from arguments concerning 
the bulk geometry based on the energy conditions that are satisfied 
or violated by the matter-content of the bulk model. In particular, 
we found a qualitative geometrical dual gravity picture demonstrating how the 
Kondo cloud acts as a screening cloud hiding the impurity. From the field 
theory side it is expected \cite{Affleck:1991tk} that due to the boundary RG 
flow, the impurity effectively looses degrees of freedom along the RG flow from 
the UV to the IR. This corresponds to the screening. In our holographic model, 
there are three clear properties that we interpret to correspond to this: First, 
the flux of the Yang-Mills field $a$ is decreasing when moving along the brane 
from the UV to the IR. This was  studied in detail in \cite{Erdmenger:2013dpa} 
for the case without backreaction and qualitatively carries over to our 
numerical solutions as well. Second, we find that due to the null energy 
condition (NEC) the renormalised volume of our bulk spacetime is reduced for 
decreasing temperature, a feature that may be interpreted as a reduced 
holographic complexity, see 
\cite{Susskind:2014rva,Susskind:2014rvaAdd,Alishahiha:2015rta,Brown:2015bva,
MIyaji:2015mia}. 
Third, the reduction of the impurity entropy both with decreasing $T$ and 
increasing $\ell$ is a clear sign of the screening of the impurity.

This was formalised in section \ref{sec::g} in form of the holographic 
$g$-theorem \cite{Affleck:1991tk,Friedan:2003yc} that 
has attracted interest in holographic research before 
\cite{Yamaguchi:2002pa,Takayanagi:2011zk,Fujita:2011fp,Nozaki:2012qd}. 
Due to the proof given in \cite{Takayanagi:2011zk}, it came as 
no surprise that we found the $g$-theorem to be satisfied by our
explicit example, specifically as a 
consequence of the NEC being satisfied by the matter 
fields in our holographic model.

In section \ref{sec::largeL} we derived an analytical approximation 
formula to our numerical results valid for large entanglement interval sizes $\ell$.
This made it possible to compare our results to similar field theory results 
that are summarised in the introduction, with remarkable 
agreement. Indeed, the generality of the construction employed in section 
\ref{sec::largeL} led us to view this as a geometrical bulk manifestation of the 
universality of perturbing about an IR fixed point, at least to first order. 
For systems that can be described by a constant tension (bulk) solution at $T=0$ 
in the AdS/BCFT approach\cite{Takayanagi:2011zk,Fujita:2011fp,Nozaki:2012qd}, 
we found that an impurity length scale ($\frac{\pi}{4}\xi_{\text{K}}$ in the 
case of our Kondo model) can be identified with a geometrical bulk length $D$ as 
defined in figure \ref{fig::fittedbrane}. In comparing gravity and field theory 
results, the significance of the
small-temperature behaviour of our model became apparent.

\subsection{Outlook}
\label{sec::outlook}

There are many open questions related to impurities in strongly
coupled systems. The results of the present paper, which provide a new
link between the impurity entropy and defects in spacetime geometry,
may be applied in a number of different scenarios. In the immediate
holographic context, we note the papers \cite{Dias:2013bwa,Horowitz:2014gva}, 
where vortices in 
holographic superfluids were considered as defects in a 2+1-dimensional CFT.  
Also in that case, the theory flows to a non-trivial fixed point in the IR: At 
low energies, the vortex corresponds to  a conformal defect interacting
with the gapless modes in the superfluid.

For Kondo or similar impurity systems, entanglement entropy is not the only 
measure of entanglement, others such as R\'enyi entropies and entanglement 
negativity (to only name a few) have been studied in 
\cite{PhysRevA.74.050305,PhysRevB.81.064429,Sodano1404,PhysRevLett.114.057203}. 
How to study these measures of entanglement in our holographic Kondo model is 
an interesting task that we leave to the future. The results of 
\cite{Sodano1404} for the two-impurity Kondo model are especially interesting  
in the light of the recent generalisation \cite{O'Bannon:2015gwa} of the 
holographic Kondo model \cite{Erdmenger:2013dpa} to the two-impurity case. 
Following the direction of research begun in \cite{O'Bannon:2015gwa}, it would 
of course also be interesting to study spatially separated impurities and Kondo 
lattices.
Moreover, it will also be interesting to analyse the case of zero temperature.
This may be compared to condensed matter results, as for instance in 
\cite{PhysRevB.81.041305,
1742-5468-2007-01-L01001,
1751-8121-42-50-504009,
1742-5468-2007-08-P08003,
PhysRevB.84.041107}.

Moreover, the Kondo effect was recently discussed in the context of QCD
\cite{Hattori:2015hka}, where it was found to occur in light quark matter
which contains heavy quarks as impurities. It will be illuminating to
compare the QCD Kondo model with the holographic one discussed here.

In view of the new recent interaction between gauge/gravity duality
and quantum information theory, we note that the Kondo model was
studied in the MERA approach for instance in \cite{KondoMERA}. Here
our model, which links the removal of points from the geometry to the
decrease of information, may also prove useful. Indeed in sections 
\ref{sec::numericalresults} and \ref{sec::summ} we pointed out that the 
screening of the impurity by the Kondo cloud might manifest itself not only 
in the entanglement or impurity entropy, but also in a quantity referred to as 
holographic complexity 
\cite{Susskind:2014rva,Susskind:2014rvaAdd,Alishahiha:2015rta,Brown:2015bva,
MIyaji:2015mia}. It~will certainly be interesting to study this in more 
detail, calculating the holographic (impurity-) complexity numerically and 
investigating its behaviour under the boundary  RG flow. One interesting 
question might be whether the holographic $g$-theorem can be phrased in terms of 
complexity instead of entanglement entropy.

\section*{Acknowledgements}
We are very grateful to
Ian Affleck, 
John Estes, 
Nabil Iqbal, 
Kristan Jensen, 
Henrik Johannesson, 
Charles Melby-Thompson,
Thore Posske,
Subir Sachdev,
Pasquale Sodano, 
Ioannis Papadimitriou, 
Jonas Probst 
and Wolfgang M\"uck 
for illuminating discussions.
We are particularly grateful to Andy O'Bannon for extremely useful comments.
JE, MF and CH thank the Galileo Galilei Institute for Theoretical 
Physics for the hospitality and the INFN for partial support during the course 
of this work.
JE, MF and JMSW are grateful to the IPMU Tokyo for hospitality.
JMSW thanks NCTS for support during the course of this work.
This work is partially supported by the Spanish grant 
MINECO-13-FPA2012-35043-C02-02. 
CH is supported by the Ramon y Cajal fellowship RYC-2012-10370. 
JE thanks the KITPC Beijing and the IIP Natal for hospitality. 
Moreover, she thanks the participants of the school `Strongly coupled field 
theories for condensed matter and quantum information theory' in Natal for 
discussions.


\appendix

\section{Junction Conditions for Abelian Chern-Simons Fields}
\label{sec::JunctionCS}

Here, we derive the junction conditions that govern the 
interaction of our (abelian) bulk Chern-Simons field with the brane 
worldvolume current. As for the Israel junction conditions 
\eqref{eq:leftoverIsrael} of the gravitational field, this can be done in 
different ways. We consider the case that the current is smoothed out as 
described below, such that after taking a well-defined limit it becomes a delta 
distribution. For a related study of the behaviour of Chern-Simons fields in the 
presence of domain walls, see \cite{Antillon:1997xr,Torres:1997be}.

\begin{figure}[htb]
 \centering
 \def\svgwidth{0.6\columnwidth}
\executeiffilenewer{current_latexed.svg}{current_latexed.pdf}%
{inkscape -z -D --file=current_latexed.svg %
--export-pdf=current_latexed.pdf --export-latex}%
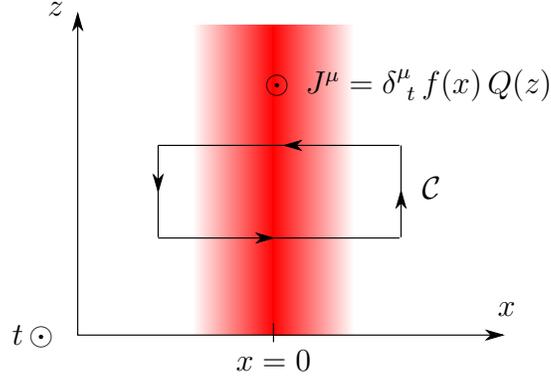%

\caption{A static worldvolume current of the brane at $x=0$, smoothed out by a 
profile $f(x)$. $\mC$ is the path of a possible 
Wilson loop.}
\label{fig::current}
\end{figure}

Later on, we will phrase our junction conditions in a manifestly gauge and 
coordinate invariant way, such that our results can be readily applied to the 
Kondo model. 
However, for now assume that we work in a Gaussian normal 
coordinate system, such that the brane is located at $x=0$. In our Kondo model, 
the current involving the charged scalar $\Phi=\phi \,e^{i\psi}$ on the brane 
worldvolume takes the 
form
\begin{align}
 J^m&=\gamma^{mn}\,i\left(\Phi\mD_n\Phi^\dagger-\Phi^\dagger\mD_n\Phi\right)
=2\,\gamma^{mn}\left(A_n-a_n+\partial_n 
\psi\right)\phi^2 \, ,
\label{current}
\end{align}
which in the static case can be written as $J^m\sim 
\mQ(z)\left(\partial_t\right)^m$ in radial gauge. The brane is now assumed 
to be smoothed out by a profile satisfying $f(x)=F'(x)$, such that there exists 
a limit in which $f(x)\rightarrow\delta(x)$, 
$F(x)\rightarrow\theta(x)$.\footnote{
E.g.~use
$f_a(x) :=  (1/\sqrt{\pi} a) \,\exp(-x^2/a^2)$ and 
$F_a(x) := (1/2) \,(1+\text{erf}(x/a))$ and take the limit $a \rightarrow 0$.
}
See figure \ref{fig::current} for the setup.

As we assume a static situation, all objects are independent of the 
coordinate $t$. With the CS equation of motion 
\begin{align}
 \epsilon^{\rho\mu\nu}F_{\mu\nu}=-4\pi \,J^\rho,
\end{align}
the only non-trivial equation of motion then reads 
\begin{align}
 2\,\epsilon^{tz}\left(\partial_z A_x - \partial_x A_z\right) =-4\pi\, J^t 
=-4\pi \, f(x)\mQ(z) \, ,
\label{CSeq}
\end{align}
where we define the worldvolume $\epsilon$-tensor by contracting the bulk 
$\epsilon$-tensor with the normal form of the brane. In Gaussian normal 
coordinates, this simply means $\epsilon^{\rho\mu x}\equiv\epsilon^{\rho\mu}$.
Equation \eqref{CSeq} can be solved in two ways:
\paragraph{Gauge $A_x=0$:}
First we assume $A_x=0$, hence
\begin{align}
 -\epsilon^{tz} \partial_x A_z &=-2\pi\, f(x)\mQ(z) \, ,
 \\
 \Rightarrow  \epsilon^{tz}A_z &=2\pi\, F(x)\mQ(z) \, .
 \label{CSGauge1}
\end{align}

In the limit $F\rightarrow\Theta,\ f\rightarrow\delta$ this means that the 
parallel component of the CS field has a discontinuity proportional to the 
charge density $\mQ$ on the brane. 

\paragraph{Gauge $A_z'=0$:}
If we assume $A_z'=0$ instead, we find
\begin{align}
 \epsilon^{tz} \partial_z A_x' & =-2\pi\, f(x)\mQ(z)
\\
\Rightarrow  -\tilde{\epsilon}^{tz}A_x'&= 2\pi\, f(x)
\int^z_0 \sqrt{-\gamma}\,\mQ(\hat{z})\,\totd\hat{z} \, ,
\label{CSGauge2}
\end{align}
where $\tilde{\epsilon}$ denotes the Levi-Civita symbol, which is related to 
the Levi-Civita tensor by 
$\epsilon^{mn} = \tilde{\epsilon}^{mn}/\sqrt{-\gamma(z)}$.
In the limit $F\rightarrow\Theta,\ f\rightarrow\delta$ we see that in this 
gauge, the component of the Chern-Simons field normal to the brane acquires a 
contribution $\sim\delta(x)$. 

Both approaches are indeed gauge-equivalent. This may be seen by explicitly 
showing that the difference of both solutions \eqref{CSGauge1} and 
\eqref{CSGauge2} is a total derivative $\totd\alpha = A' - A$ with
\begin{align}
 \alpha =\frac{2\pi}{\tilde{\epsilon}^{tz}} F(x)\int^z_0 
\sqrt{-\gamma}\,\mQ(\hat{z})\,\totd\hat{z}\, .
\end{align}
In the limit in which the current distribution becomes infinitely thin, this 
means that the gauge function $\alpha$ approaches the Heaviside $\theta$ 
function, and hence $\alpha$ need not vanish for large $|x|$. Moreover, Wilson 
lines transform in the appropriate way 
\begin{align}
 W(a,b)\rightarrow e^{i\alpha(a)}W(a,b)e^{-i\alpha(b)}.
\end{align}
For a simple rectangular path such as $\mC$ in figure \ref{fig::current}, we 
may explicitly calculate the Wilson loop in the two gauges presented above and 
check that the results agree. By the equations of motion of the CS field, we 
find
\begin{align}
 \oint_{\mC} A = \int_{\text{int}(\mC)} dA \stackrel{EOMs}{\propto} 
\int_{\text{int}(\mC)} \mQ \, ,
\end{align}
where $\text{int}(\mC)$ denotes the interiour of $\mC$.

What we have learned above is that away from the brane on which the current 
$J^\mu$ is localised, the CS field satisfies its bulk (vacuum) equations of 
motion, while at the brane, depending on a gauge choice, we either have a 
discontinuity in the parallel component of the $A$-field or a $\delta$-peak in 
the normal component. 
In the former case, the projection of the CS field to the brane 
worldvolume is not uniquely defined. For symmetry reasons, we assume 
that in equations \eqref{covDer}, \eqref{current} the scalar field $\Phi$ 
couples 
to the \textit{mean value} of the CS fields projected onto the brane from the 
left (with projector $\overrightarrow{\mP}$) and from the right (with 
projector $\overleftarrow{\mP}$) side of the bulk: 
\begin{align}
 A_m\equiv\frac{1}{2} \left(\overleftarrow{\mP}_{\!\!m}^{\ \mu} A_{\mu}
+\overrightarrow{\mP}_{\!\!m}^{\ \mu} A_{\mu}\right) \, .
\label{formerD}
\end{align}
In addition, we define the discontinuity of the parallel components at the 
brane by 
\begin{align}
 C_m\equiv \overleftarrow{\mP}_{\!\!m}^{\ \mu} A_{\mu}
-\overrightarrow{\mP}_{\!\!m}^{\ \mu} 
A_{\mu}
\end{align}
and the $\delta$-contribution to the normal component of the CS field by 
\begin{align}
A^0(z,t)\equiv \int_{-\epsilon}^{\epsilon} A^{\mu} n_{\mu} \,\totd s\, ,
\end{align}
where $n_{\mu}$ is the normal form of the brane and the integral is along an 
integral curve of the vector field $n^\mu$ that crosses the brane at 
coordinates $(z,t)$. Taking the above results and definitions together, we then 
find that at the location of the brane, the CS field satisfies the junction 
condition
\begin{align}
-2\pi\, J^m=\epsilon^{im}\left(C_i-\partial_i A^0\right).
\label{CSjunction}
\end{align}
Let us comment on the interesting properties of this equation. First of all, 
we note that in the equations of motion of the scalar field and the 
Yang-Mills field, as well as in the current \eqref{current}, only the 
projection 
$A_m$ as defined in \eqref{formerD} plays a role, while the quantities $C_m$ 
and $A^0$ appear only on the right hand side of \eqref{CSjunction}. 
Furthermore, for given $J^m$ equation \eqref{CSjunction} is an algebraic 
equation for $C_m$ and $A^0$. This means that just as in 
\cite{Erdmenger:2013dpa} the CS field decouples from the dynamics of the brane 
and we can first gauge $A_m=0$ in our setup, then solve for $\Phi,a_m$ and 
$J^m$ and then solve the algebraic equation \eqref{CSjunction}. Finally, it is 
interesting to note the form of \eqref{CSjunction} in which $J^m$ and hence the 
combination $\left(C_i-\partial_i A^0\right)$ are gauge 
invariant by definition, such that the quantity $A^0$ acts as a gauge function 
for $C_m$.    

\section{Details of Numerics}
\label{sec::NumericsApp}

Here we briefly describe our method to solve the equations of 
motion \eqref{eq:EOMscalar}-\eqref{eq:EOMemb} numerically. First of all we 
exploit two scaling symmetries of the EOMs to set $L=1$ and 
$z_H=1$. All coordinates become dimensionless quantities: $\tilde{z} = z/z_H, 
\tilde{x} = x/z_H, \tilde{t}=t/z_H$. The relationship between the matter 
action and the gravity action is fixed by $\kappa_N^2  \mN = 1$. 
The electric flux defined in section \ref{sec::EOMs} is set to $\mC = 1/2$ to 
be consistent with \cite{Erdmenger:2013dpa}.

The $\tilde{z}\tilde{z}$-component of \eqref{eq:EOMemb} only contains first 
derivatives of the 
embedding function $\tilde{x}_{+}(\tilde{z})$. It can be analytically solved 
for 
$\tilde{x}_{+}'(\tilde{z})$ as a function of $\phi$ and $a_t$ as well as their 
derivatives. 
There are four solutions to this equation. However, only one of them is 
consistent 
with the background solution \eqref{backgroundSolutionGaugeAndEmbedding}. We 
take this one and insert it into the equations of motion for $\phi$ and $a_t$.
These are now two coupled second order ODEs for two fields and hence require 
four boundary conditions.

To solve them numerically, we apply the shooting method. We expand the 
fields at the asymptotic boundary ($\tilde{z}=0$, UV) and at the horizon 
($\tilde{z}=1$, i.e.~$z = z_H$, IR). 
Since the mass of the scalar is fine-tuned to saturate the 
Breitenlohner-Freedman bound \eqref{BFbound}, the UV expansions come along 
with roughly $\mathcal{O}(1)+3 k$ logarithms at $k^{\text{th}}$ order 
in $\tilde{z}$:
\begin{align}
\phi(\tilde{z}) & \sim \sqrt{\tilde{z}} \left( \alpha_T \log(\tilde{z}) 
	+ \beta_T + \sum_{k=1}^\infty \sum_{j=0}^{3 k + 3} 
	\left(c_{\phi}(k,j)\, \tilde{z}^k\, \log(\tilde{z})^j \right) \right)\\
a_t(\tilde{z})  & \sim  Q \frac{z_H}{\tilde{z}} + \mu_T 
    + \sum_{k=1}^\infty 
      \sum_{j=0}^{3 k + 4} \left( c_{a}(k,j) \,\tilde{z}^k\, \log(\tilde{z})^j 
\right)
\end{align}
For this reason, expanding the fields in the UV using \textsc{Mathematica} 
takes remarkably long and we were able to expand the fields only up to 
$4^{\text{th}}$ order in the UV. 
In the IR, we expand to $8^{\text{th}}$ order. $Q$ is fixed by 
\eqref{eq:expansionAfirstorder} and our choice of $\kappa_N^2  \mN$ and $\mC$. 
Regularity at the horizon fixes two more integration constants such that we are 
left with a one-parameter family of solutions, which we label by the 
value of the chemical potential $\mu_T = \mu / 2\pi\, T_c$. 
Starting at the critical value $\mu_T = \mu_{T_c}$, see below 
\eqref{eq:expansionAfirstorder}, we choose a set of increasing values 
of $\mu_T$. 
For each of those we vary $\alpha_T$ and $\beta_T$ and integrate the EOMs 
numerically from 
the boundary to the horizon until the regularity conditions 
\eqref{horizonconsistency} are satisfied. Hence, each solution can also be 
referred to by $\kappa_T$ and, by \eqref{eq:ToverTK_vs_kappaT} and 
\eqref{eq:Tc}, a value of $T/T_c$.

Next, we define a set of radial positions between the conformal boundary and 
the horizon.
For each position we obtain the embedding profiles $\tilde{x}_+(\tilde{z})$ by 
integrating numerically 
$\tilde{x}_+'(\tilde{z})$ as a function of $\phi$ and $a_t$, using the 
numerical solutions 
found by the shooting. We choose the integration constant such that 
$\tilde{x}_+(0) = 
0$, which is possible due to translational symmetry of the BTZ metric 
\eqref{BTZlinelement}.

We obtain numerical data in the form of $(\tilde{z}, \tilde{x}_+(\tilde{z}), 
\tilde{x}_+'(\tilde{z}))$ for different 
$\tilde{z}$, which can be mapped to a unique geodesic starting normal to the 
brane at 
$(\tilde{z},\tilde{x}_+(\tilde{z}))$ and ending at the asymptotic boundary at 
$\tilde{z}= 0$ and $\tilde{x} = \tilde{\ell} $.  
The geodesics are regularised by subtracting the divergent term 
$\log(2/\tilde{z})$ as 
we take the limit $\tilde{z} \rightarrow 0$ and have regularised 
(dimensionless) length $\mL(\tilde{\ell})/L$, where we reintroduced the AdS 
radius $L$ appearing in the line element. 

Thus, starting at one position $(\tilde{z},\tilde{x}_+(\tilde{z}))$ on the 
brane, we obtain data of the form $(\mL(\tilde{\ell})/L,\tilde{\ell})$.
Repeating this procedure for many starting points on the brane and for 
different $\mu_T$, i.e.~$T/T_c$, we can interpolate a function 
$\mL(\tilde{\ell})/L$. Combined with \eqref{RT}, we obtain the entanglement 
entropy as a function of the subinterval around the impurity and the 
temperature. 
This is how we compute the first term in the definition of the impurity entropy 
\eqref{Simpdef}. The second term is known analytically \eqref{BTZresult}.


\providecommand{\href}[2]{#2}\begingroup\raggedright\endgroup

 
\end{document}